\documentclass[acmsmall]{acmart}

\AtBeginDocument{%
  \providecommand\BibTeX{{%
    \normalfont B\kern-0.5em{\scshape i\kern-0.25em b}\kern-0.8em\TeX}}}

\setcopyright{acmcopyright}
\copyrightyear{2020}
\acmYear{2020}
\acmDOI{10.1145/3408890}

\acmJournal{TIST}
\acmVolume{11}
\acmNumber{6}
\acmArticle{69}
\acmMonth{9}

\usepackage{amsthm} 
\newtheorem{thm}{Theorem}
\theoremstyle{remark}
\newtheorem{defn}{Definition}[section]
\newtheorem{exmp}{Example}[section]
\usepackage[ruled,linesnumbered]{algorithm2e}
\SetKwComment{Comment}{$\triangleright$\ }{}
\usepackage{multirow,longtable,fancyhdr,figsize}
\usepackage{tabularx}
\usepackage{graphicx}
\usepackage{subfigure}
\usepackage{bm}
\usepackage{url}
\usepackage{makecell}
\usepackage{xcolor}
\usepackage{hyperref}
\usepackage{pbox}
\usepackage[color]{changebar}
\cbcolor{blue}
\nochangebars
\hypersetup{
    colorlinks=true,
    linkcolor=blue,
    filecolor=magenta,      
    urlcolor=cyan,
}

\begin{document}

\title[BOXREC]{BOXREC:   Recommending a Box of Preferred Outfits in Online Shopping}
\author{Debopriyo Banerjee}
\email{debopriyo@iitkgp.ac.in}
\orcid{0000-0001-9773-776X}
\author{Krothapalli Sreenivasa Rao}
\email{ksrao@cse.iitkgp.ac.in}
\author{Shamik Sural}
\email{shamik@cse.iitkgp.ac.in}
\orcid{0000-0002-4315-7329}
\author{Niloy Ganguly}
\email{niloy@cse.iitkgp.ac.in}

\affiliation{%
\department{Dept. of Computer Science and Engineering}
  \institution{Indian Institute of Technology Kharagpur}
  \state{West Bengal}
  \postcode{721302}
  \country{India}
}

\renewcommand{\shortauthors}{D. Banerjee, et al.}

\begin{abstract}
Fashionable outfits are generally created by expert fashionistas, who use their creativity and in-depth understanding of fashion to make attractive outfits. Over the past few years, automation of outfit composition has gained much attention from the research community.
\cbstart
Most of the existing outfit recommendation systems focus on pairwise item compatibility prediction (using visual and text features) to score an outfit combination having several items, followed by recommendation of top-n outfits or a capsule wardrobe having a collection of outfits based on user's fashion taste.
\cbend
However, none of these consider user's preference of price-range for individual clothing types or an overall shopping budget for a set of items.
\cbstart
In this paper, we propose a box recommendation framework~-~BOXREC~-~which at first, collects user preferences across different item types (namely, top-wear, bottom-wear and foot-wear) including price-range of each type and a maximum shopping budget for a particular shopping session. It then generates a set of preferred outfits by 
retrieving all types of preferred items from the database (according to user specified preferences including price-ranges),
creates all possible combinations of three preferred items (belonging to  distinct item types) and
verifies each combination using an outfit scoring framework~-~BOXREC-OSF. Finally, it 
provides a box full of fashion items, such that different combinations of the items maximize the number of outfits suitable for an occasion while satisfying maximum shopping budget. 
We create an extensively annotated dataset of male
fashion items across various types and categories (each having associated price) and a manually annotated positive and negative formal as well as casual outfit dataset.
\cbend
\cbstart
We consider a set of recently published pairwise compatibility prediction methods as competitors of BOXREC-OSF.
Empirical results show superior performance of BOXREC-OSF over the baseline methods.
We found encouraging results by performing both quantitative and qualitative analysis of the recommendations produced by BOXREC. Finally, based on user feedback corresponding to the recommendations given by BOXREC, we show that disliked or unpopular items can be a part of attractive outfits.
\cbend
\end{abstract}

\ccsdesc[300]{Computing methodologies~Neural networks}
\ccsdesc[300]{Computing methodologies~Learning latent representations}
\ccsdesc[300]{Information systems~Recommender systems}
\ccsdesc[300]{Theory of computation~Problems, reductions and completeness}

\keywords{E-commerce, fashion, outfit compatibility, neural networks, optimization, pagination, composite recommendation}
\maketitle

\section{Introduction}
In the next few years, the global fashion industry is expected to grow in value from 1.3 trillion to 1.5 trillion US dollars\footnote{\url{https://www.statista.com/topics/5091/apparel-market-worldwide/}}. 
The ultimate driving force behind this enormous growth is the proliferation of online shopping portals, which has facilitated access to vast collections of fashionable apparels to millions of people globally.
Hence, efficient techniques such as clothing modeling, recognition, parsing, retrieval, analysis and recommendations are required to provide the best shopping experience.

Generally, recommender systems form the backbone of any shopping site that help users to navigate the most relevant list of single items.
Deviating from this conventional idea, Xie et al. \cite{Xie:2010} introduce a new composite recommender system that assists a user to buy a package or box full of interrelated set of items.
In line with this concept, recently, websites like Outfittery\footnote{\url{https://www.outfittery.com/}}, StyleCraker\footnote{\url{https://www.stylecracker.com/}}, etc., 
\cbstart
have introduced the concept of shopping a set of items from different item types based on user-specified preferences, body measurements, occasion, price-range for individual clothing types and an estimated total shopping budget. On placing an order, they ship a surprise box filled with fashion items that are manually chosen by their style experts and provides suggestions for assembling different outfits from the shipped items. 
This problem requires retrieval of a minimum number of preferred items (satisfying type-specific price-ranges and the total shopping budget) that mix and match well with each other, i.e., diverse combinations help in making compatible outfits. It has been partially addressed in some of the recent works \cite{Hsiao:2018, Dong:2019:PCW} that focus on
automating the creation of capsule wardrobes. 

Hsiao et al. \cite{Hsiao:2018} formulate capsule creation as the problem of selecting a subset of items having minimum clothes and accessories that jointly maximizes versatility and compatibility between the items. Dong et al. \cite{Dong:2019:PCW} consider an existing wardrobe of a user and creates a personalized capsule wardrobe (PCW) adding and deleting some items. PCW jointly incorporates user-garment modeling and garment-garment modeling, where user body shape factor has been included in addition to the user preference factor in the user-garment modeling. But none of the above methods consider type-specific price-ranges and total shopping budget. 

As stated by Zheng et al. \cite{Zheng:2020:PriceawareRGCN}, item price plays a vital role in addition to item preference in determining final purchase decision. E.g., a user may be interested in a particular top-wear (having price in the range $1$K - $3$K), because of its attractive visual appearance. But the user normally spends $<$ $1$K for top-wear items. In such a case, the recommender system should suggest a visually similar item whose price is $<$ $1$K, otherwise the user would not initiate purchase. Moreover, there is a maximum amount that a user can spend when shopping for a capsule wardrobe. Hence, it is also essential to consider the user's maximum budget for a particular shopping session on any e-commerce portal.
\cbstart
\begin{figure}[t]
	\centering
	\includegraphics[trim=0 0 0 0,clip,width=\textwidth]{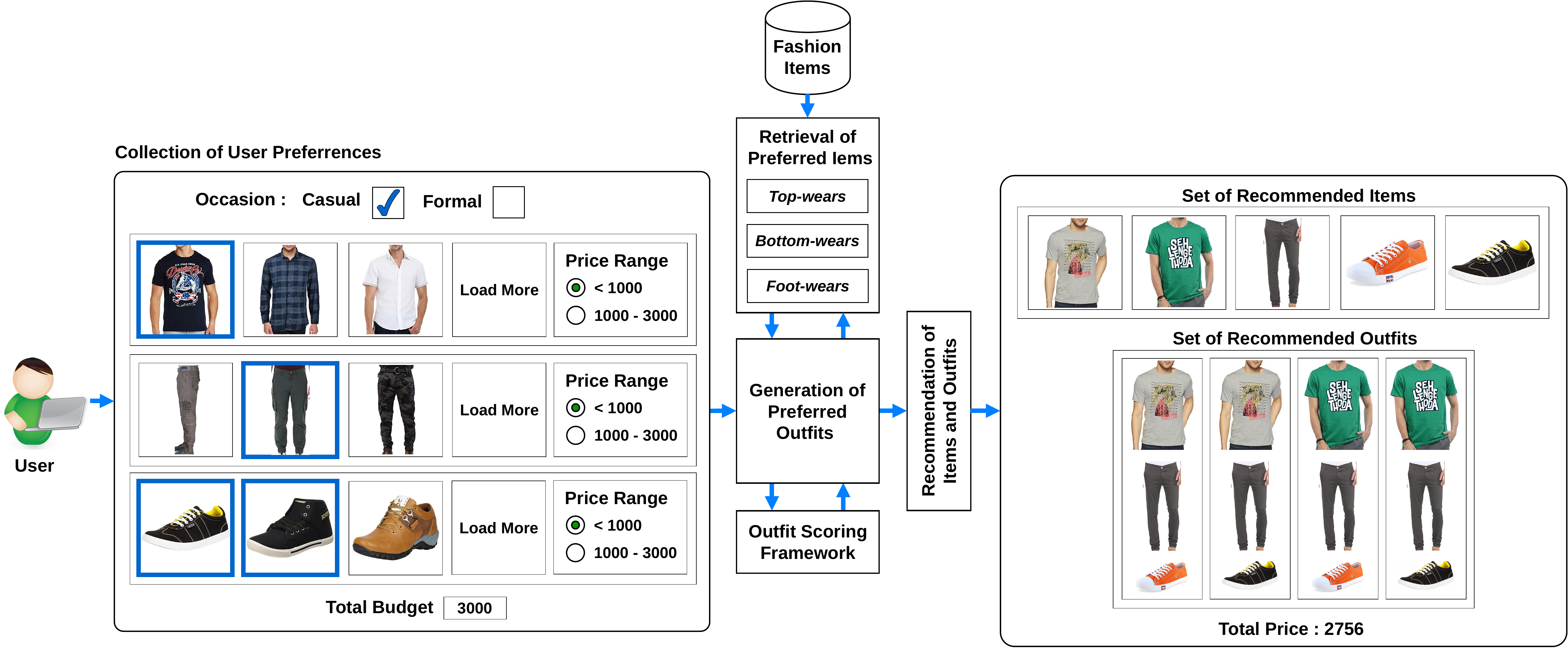}
	\caption{Overview of BOXREC: User first mentions the Occasion (e.g., casual), then the set of Preferred Items from top-wears, bottom-wears and foot-wears (highlighted in blue), followed by Price-Range of each type (e.g., $< 1000$) and a Total Budget (e.g., $3000$). BOXREC then generates a set of Preferred Outfits by first Retrieving a set of Preferred Items (from each item type), creates all possible combinations (each having three items, one from each type), verifies the quality of the outfits using an Outfit Scoring Framework and finally Recommends a set of Preferred Items that compose maximum possible Outfits (satisfying the Total Budget).}
	\label{fig:overview_of_boxrec}
\end{figure}
\cbend
This motivated us to propose a box recommendation (BOXREC) framework considering users's fashion preferences including type-specific price-ranges and an overall budget. 

As shown in Fig. \ref{fig:overview_of_boxrec},   BOXREC first collects user preferences of top-wear, bottom-wear and foot-wear items that are either casual or formal, including respective price-ranges and a maximum total shopping budget. The second step is generation of preferred outfits, which involves retrieval of similar top-wear, bottom-wear and foot-wear items from the database of all fashion items, according to the mentioned preferences and price-ranges of each item type, 
followed by creation of all possible combinations of three items (i.e., one item from each type) using the set of retrieved items. But all of the combinations may not qualify as a valid outfit. So, we employ an outfit scoring framework (BOXREC-OSF) that iteratively verifies each of the combinations and helps in generating a set of preferred outfits. Finally, BOXREC solves an optimization problem: box recommendation with maximum outfits (BOXRECMO) that packs maximum number of preferred outfits in a box, where total price of all the items in the recommended box satisfies user mentioned overall budget.
However, a user may not like certain items and outfit combinations in the box. BOXREC records both positive and negative user feedback against items and outfits for further analysis. 
\cbend
The main contributions of our work are as follows: 
\cbstart
\begin{itemize}
    \item We present a novel box recommendation~-~BOXREC framework that makes price aware recommendation of a box filled with preferred fashion items, which in turn maximizes the number of corresponding preferred outfit combinations, satisfying a maximum shopping budget.
    \item For generating preferred outfits, we develop an outfit scoring framework~-~BOXREC-OSF that is powered by an enhanced image encoder (better than \cite{Lin:2019}) and exploits multi-modal features (visual and textual) for computing compatibility between preferred item pairs. We also propose an efficient aggregation scheme for aggregating the pairwise compatibility scores in an outfit.
    \item The recommendation part of BOXREC is an optimization problem. We formally define the box recommendation with maximum outfits~-~BOXRECMO problem, prove that it is NP-Complete and use a heuristic solution to solve the optimization objective.
    \item We construct a new male fashion dataset having items from different types and categories, a male outfit dataset annotated by fashion experts and a user-item preference dataset (including type-specific preferred price-ranges and an estimated total budget for a particular shopping session).
\end{itemize}
\cbend
The rest of the article is structured as follows: Section \ref{RelatedWork} briefly surveys related work on outfit recommendation and composite package recommendation systems, respectively. Section \ref{notations_and_prob_formulation} introduces relevant notations, problem formulation and the working of the BOXREC framework in detail. Section \ref{heuristics} presents a heuristic to solve the optimization problem of BOXRECMO. Experiments and comparisons are provided in Section \ref{exp_eval}. Implementation of the BOXREC framework as a web application follows in Section \ref{web_app}. Finally, we conclude the article in Section \ref{conclusion}.

\section{Related Work}\label{RelatedWork}
Clothing fashion demonstrates the aesthetic appreciation of outfits and reflects the development status of our society.
In recent years, due to the proliferation of online shopping portals, vast collections of fashionable apparels are easily accessible. However, users are overwhelmed with the number of available options and find it difficult to scan through the entire stock of catalogued items.
\cbstart
This necessitates the presence of recommender systems for assisting users to navigate the most relevant items.
In the fashion domain, there are several interesting contributions on street to shop \cite{Liu:2012:Street2Shop, Kiapour:2015:WhereToBuyIt},
body shape \cite{Hidayati:2018:PersonalBodyShape,Hsiao:2019:ViBE}, price and popularity \cite{Nguyen:2014:LTRPFRSIF} based recommender systems, where fashion items are generally represented by visual \cite{Yuhang:2016, KedanLi:2019} or textual features \cite{Hsiao:2018} or a combination of both \cite{YunchengLi:2017,Han:2017, Vasileva:2018:LTAE,Dong:2019:PCW}. 
Recently, price has been shown to be an important factor \cite{Zheng:2020:PriceawareRGCN} in modeling user's purchase intention as part of a recommender system. In this work, we consider  both visual and textual features to represent fashion images and consider user preferences including preferred price-ranges across different clothing types to retrieve relevant items. 

Apart from solving the well known fashion item retrieval problem, currently, researchers from both academia and industry are working towards developing efficient outfit recommender systems
\cite{Tangseng:2017, YunchengLi:2017, Han:2017, Song:2017, Hsiao:2018, He:FashionNet:2018, Vasileva:2018:LTAE, KedanLi:2019, Lin:2019, Chen:2019:POG,Song:2019:CompModelBook, Han:2020:NCMPKD}, each of which employs an outfit scoring function to verify or compose outfit combinations that promote sale of a set of items from different
categories that constitute an outfit. 
The most simple outfit scoring function concatenates the CNN features extracted from a fixed set of apparel items and predicts whether the combination is good or bad \cite{Tangseng:2017}. This naive method can be further improved by learning separate matching networks for distinct clothing pair types \cite{He:FashionNet:2018, Vasileva:2018:LTAE, KedanLi:2019, Lin:2019}, and outfit compatibility can be determined by averaging the scores of all item pairs in an outfit. The discriminative method of Li et al. \cite{KedanLi:2019} is shown to perform better than the type aware compatibility prediction framework \cite{Vasileva:2018:LTAE}.

Furthermore,
compatibility between item pairs depend on whether an item complements particular features of another item and vice versa. As a result,
mutual attention between fashion image pairs \cite{Lin:2019}
better models the matching relation between them. However, such a method requires suitable representation of fashion product images. We employ mutual attention in our outfit scoring framework, but replace the image encoder used in \cite{Lin:2019} with an enhanced image encoder, based on Inception V3 architecture (pre-trained on fashion images) for extracting intermediate convolutional feature maps of input fashion images. As a side information, Lin et al. augments visual features with latent factor based item features. But, such an approach has a drawback. It requires the presence of same set of items in training and test data. So, we use textual features (including category and product title/description) in place of latent factor features.

Compatibility relationship between items in an outfit can also be learned using a bi-directional LSTM that predicts the next item conditioned on the previous ones \cite{Han:2017}, where an outfit is considered as a sequence of items from top to bottom, followed by accessories, where each item acts as a time step (similar to words in a sentence). This method is capable of recommending an item that matches existing components in a set to form a stylish outfit and generating an outfit with multi-modal (image/text) specification provided by a user. However, unlike words, which have a particular order of appearance in a sentence, items in an outfit do not have any order. So, modeling outfits as an ordered sequence is practically not correct. 

Another way of outfit generation is incrementally adding items to a set, and every time checking the quality of the current set of items using an outfit quality scorer \cite{YunchengLi:2017}. Instead of incrementally adding items, we first create different combinations of fashion items, each having three items (i.e., one top-wear, bottom-wear and foot-wear), and then verify the compatibility of the items. We shortlist the verified compatible outfit combinations and use them during recommendation phase.

In practice, a classical recommender system gives a list of recommendations where each recommendation is a single item \cite{Liu:2013:STC,Cheng:2016:UPR,Golubtsov:2016,Zhao:2019:PRG}.
\cbend
For instance,
Golubtsov et al. \cite{Golubtsov:2016} conduct a survey to collect a user's preferred color type, age, price-range and preference of items (like / dislike).
\cbstart
They use this data to propose a ladies-wear recommender system based on reinforcement learning that suggests top-k relevant single items to users. Instead of single items, a recommender system can also suggest a package or a box full of items \cite{Xie:2010} based on user mentioned maximum total budget, where each item can have an associated value (rating) and a cost. 

In the context of fashion, package or composite recommendation corresponds to recommendation of capsule wardrobes \cite{Hsiao:2018, Dong:2019:PCW}. Capsule wardrobes are either constructed from scratch by including a set of inter-compatible items to provide maximal mix-and-match outfits \cite{Hsiao:2018}, or an existing wardrobe of a user is updated by adding and deleting some items \cite{Dong:2019:PCW}. But none of them consider user's price-range (type-specific) preferences and an overall shopping budget \cite{Xie:2010}, which are important factors for making final purchase decision. 

In this paper, we consider user's item preferences including price-range across different clothing types and an overall budget (for a particular shopping session) for recommending a box full of items that maximize the number of possible outfit combinations. As mentioned before, we incorporate suitable enhancements over the mutual attention mechanism of Lin et al. Our proposed modifications help in making better predictions of pairwise compatibility and outfit scoring. We also introduce a new aggregation scheme for aggregating pairwise compatibility scores in an outfit based on logical AND operation that makes less false positive and false negative predictions compared to the usual summation/average scheme.
\cbend 


\section{Notations and Problem Formulation}\label{notations_and_prob_formulation}
We first explain the notations used in this paper.
Let $\mathcal{T}$ be the set of all clothing types (e.g., top-wear ($tw$), bottom-wear ($bw$), foot-wear ($fw$), etc.) and $\mathcal{I}$ = $I_1 \cup I_2 \cup\ldots\cup I_{|\mathcal{T}|}$ be the set of all fashion items, where each $I_t$ = \{$x_{t,1}, x_{t,2},\ldots,x_{t,|I_t|}$\} denotes the set of items of type $t$. In general, $x$ denotes a single item of $\mathcal{I}$, $x_t$ is an item of $I_t$ and $x_{t,i}$ is the $i$-th item of $I_t$. Each item $x_{t,i}$ is associated with two modalities such as image and category.
Let $x^k_{t,i}$ denote the $k$-th modality of item $x_{t,i}$, so that $x_{t,i}$ = \{$x^1_{t,i}, x^2_{t,2},\ldots,x^\mathcal{K}_{t,i}$\}, where $\mathcal{K}$ is the total number of modalities\footnote{\cbstart We have considered $\mathcal{K}=3$:~i) \textit{image} ($x^1_t$) ii) \textit{category} ($x^2_t$) and iii) \textit{product title/description} ($x^3_t$), as the three modalities.
\cbend
}. 
Next, we define an outfit as follows.
\begin{defn}\textit{Outfit~-~}\cbstart
A set of items which has exactly one element from each of top-wear, bottom-wear and foot-wear that makes a person suitably presentable for an occasion.

In general, we denote an outfit as a set $o$ having one item from each set $I_t\in \mathcal{I} (\forall t\in \mathcal{T})$\footnote{In this paper, we assume $\mathcal{T} = \{tw,\ bw,\ fw\}$.}. A sample casual outfit, for example, may consist of a tshirt (top-wear), jeans (bottom-wear) and sneakers (foot-wear). 
\end{defn} 
\cbend
\cbstart
We define a list of notations
for items and outfits (generally denoted as $x$ and $o$, respectively).
\cbend
Let $H$ be a set of outfits. Then:
\begin{itemize}
\item the \textbf{size} $|o|$ of $o$ is the number of items in $o$.
\item the \textbf{price of item} $x$ is $price(x)$. 
\item the \textbf{occasion} for item $x$ is $occ(x)$.
\item the \textbf{price of outfit} $o$ is the sum of the prices of all the items in $o$: $price(o)$ = $\sum_{x\in o} price(x)$.
\item the \textbf{size} $|H|$ of $H$ is the number of outfits in $H$.
\item the \textbf{cardinality} $Card(H)$ is the total number of items (distinct or not) in $H$: $Card(H)$ = $\sum_{o\in H}|o|$.
\item the \textbf{multiplicity} of item $x$ in $H$, $\mu_H(x)$ is the number of occurrences of $x$ in the outfits of $H$: $\mu_{H}(x)$ = $|\{o\in H : x \in o\}|$.
\item the \textbf{relative size} of $o$ in $H$ is the sum of the reciprocals of the multiplicities of the items of $o$ in $H$: $|o|_H$ = $\sum_{x\in o}\frac{1}{\mu_H(x)}$.
\item the set of \textbf{distinct items} $\nu(H)$ is the set of distinct items present in the union of all the outfits in $H$: $\nu(H) = \cup_{o \in H}o$. 
\item the \textbf{total price}  $T(H)$ is the sum of the prices of all the items in $\nu(H)$: $T(H) = \sum_{x\in \nu(H)}price(x)$. 
\end{itemize} 
We propose a multi-item box recommendation (BOXREC) framework (Fig. \ref{fig:overview_of_boxrec}) that helps a user to shop for a collection of items such that different combinations of the items maximize the corresponding number of outfits suitable for an occasion, satisfying user's type-specific fashion taste, price-range preference and an overall budget constraint. In Sub-sections \ref{collection_of_user_pref_gen_pref_outfits}-\ref{reco_step}, we explain in detail about each of the steps of BOXREC depicted in Fig. \ref{fig:overview_of_boxrec}. 
\begin{algorithm}[t]
    \SetAlgoNoLine
    \SetArgSty{textnormal}
    \DontPrintSemicolon
	\KwIn{Set of all items $\mathcal{I} = \cup_{t\in\mathcal{T}} I_t$, an occasion $oc$, set of chosen items $I'_t$, price-range $[b^l_t, b^u_t)$ of each item, number of preferred items $m_t$ for each type $t\in \mathcal{T}$ and required number of preferred outfits $L$.} 
	\KwOut{Set of preferred outfits $O^p$, such that $|O^p| = L$. }
	\Begin{
	    $O^p \leftarrow \{\}$\\
	    \While{\cbstart $true$ \cbend}{
	         $\mathcal{I}^p \leftarrow \{ \}$\\ \For(\Comment*[f]{\normalfont{iterate over all item types}}){\normalfont{\textbf{each} type} $t \in \mathcal{T}$}{
	            $I^p_t \leftarrow RPI(I_t, I'_t, oc, b^l_t, b^u_t, m_t)$\Comment*[r]{\normalfont{retrieve preferred items of type} $t$}
	           $\mathcal{I}^p \leftarrow \mathcal{I}^p \cup I^p_t$\Comment*[r]{\normalfont{update} $\mathcal{I}^p$}
	         }
	        \cbstart
    		$O \leftarrow$ all possible combination of $|\mathcal{T}|$ items, each with one item from every $I^p_t \in \mathcal{I}^p$
    		\cbend
    		
        	\For(\Comment*[f]{\normalfont{iterate over all outfits}}){\normalfont{\textbf{each} outfit} $o_j \in {O}$}{
    			\uIf(\Comment*[f]{\normalfont{\cbstart
    			check compatibility score}
    			\cbend}){$\psi(o_j;\theta)= 1$}{
    				$O^p \leftarrow  O^p \cup \{o_j\}$\Comment*[r]{\normalfont{update $O^p$ with a new outfit}}  
    			    \cbstart	\uIf(\Comment*[f]{\normalfont{check whether size of $O^p$ equals $L$}}){$|O^p| = L$}{
    			    \textbf{return} $O^p$\Comment*[r]{\normalfont{return the set of preferred outfits}}
    			    }
    			    \cbend
    			}
    		}
    		\cbstart
    		\For(){\normalfont{\textbf{each}    type} $t \in \mathcal{T}$}{
    		$I_t \leftarrow I_t \setminus I^p_t$\Comment*[r]{\normalfont{remove all the items of $I^p_t$ from $I_t$}}
    		}
    		\cbend
		}
	}
	\caption{Generation of Preferred Outfits}
	\label{alg:ret_outfits}
\end{algorithm}

\cbstart
\subsection{Collection of User Preferences and Generation of Preferred Outfits}\label{collection_of_user_pref_gen_pref_outfits}
As shown in Fig. \ref{fig:overview_of_boxrec}, BOXREC first collects user preferences across different clothing types for a particular occasion (formal/casual), including the preferred price-range of each type and a maximum total shopping budget. 
The next step of BOXREC is generation of preferred outfits by using Algorithm \ref{alg:ret_outfits}, where Algorithm \ref{alg:ret_items} (Sub-section \ref{ret_pref_items}) is used as a sub-routine for retrieving preferred items of each type $t \in \mathcal{T}$ and an outfit scoring function (Sub-section \ref{outfit_scoring_framework}) verifies compatibility between items in each combination.

Given the set of all items $\mathcal{I} = \cup_{t\in\mathcal{T}} I_t$, an occasion $oc$, a set of chosen items $I'_t$, price-range $[b^l_t, b^u_t)$ of each item and the number of preferred items $m_t$ for each type $t\in \mathcal{T}$ (where $m_{tw}=15$, $m_{bw}=3$ and $m_{fw}=2$), Algorithm \ref{alg:ret_outfits} returns a set $O^p$ having $L$ preferred outfits. 
As shown in the algorithm,  
$O^p$ is initialized to an empty set (Line 2).
Next, for each item type $t\in \mathcal{T}$, a set of preferred items $I^p_t$ is retrieved by calling the procedure $RPI(I_t, I'_t, oc, b^l_t, b^u_t, m_t)$ of Algorithm \ref{alg:ret_items} (Line 6) and included in $\mathcal{I}^p$ (Line 7). 
In Line 9, $O$ is assigned with all possible combinations of items from $|\mathcal{T}|$ sets, where a combination has one item from each set $I^p_t \in \mathcal{I}^p$. 

Now, for each outfit $o_j\in O$, the algorithm checks at Line 11, whether $\psi(o_j;\theta)=1$ (using the outfit scoring framework explained in Sub-section \ref{outfit_scoring_framework}),
in which case $o_j$ is included in $O^p$ (Line 12).
When $|O^p|$ equals $L$ (Line 13), the algorithm terminates by returning the set $O^p$ (Line 14). At the end of each iteration of the main loop (Lines 3 to 19), the items of each $I^p_t$ are removed from $I_t$ (Line 17) so that for more than one iteration of the main loop, a new set of preferred items $I^p_t$ gets retrieved at Line 6. 

Now, suppose, the input occasion as $casual$ and price-range $< 1$K for $tw$, $bw$ and $fw$, respectively. Let, $m_{tw} = 15$, $m_{bw}=3$, $m_{fw}=2$ and $L=90$.  After retrieving $\mathcal{I}^p$, during first iteration of the main loop (Algorithm \ref{alg:ret_outfits}), we have $|O| = I^p_{tw} \times I^p_{bw} \times I^p_{fw} = 90$. Let, out of the $90$ combinations in $O$, $60$ are verified as good outfits, i.e., currently $|O^p|=60 (\neq L)$. Next, $I^p_{tw}$, $I^p_{bw}$ and $I^p_{fw}$ are removed from $I_{tw}$, $I_{bw}$ and $I_{fw}$, respectively. During the second iteration, a new set of $\mathcal{I}^p$ is retrieved, $O$ is re-initialized with $90$ new combinations; $|O^p|$ equals $L$ after checking $50$ combinations from $O$, in which case the algorithm terminates by returning $O^p$.  
\cbend
\begin{algorithm}[t]
\caption{
Retrieval of Preferred Items
}\label{alg:ret_items}
\SetAlgoNoLine
\SetArgSty{textnormal}
\DontPrintSemicolon
    \textbf{procedure} RPI($I_t$, $I'_t$, $oc$, $b^l_t$, $b^u_t$, $m_t$)\\
	    \Indp $\mathcal{C} \leftarrow$ set of categories of the items in $I'_t$\\  
	    $I^p_t \leftarrow$ \{ \}\Comment*[r]{$I^p_t$ \normalfont{will finally contain} $m_t$ \normalfont{preferred items}}
	    \For(\Comment*[f]{\normalfont{iterate over all items}}){\textbf{each} item $x_{t,i} \in I_t$}{
	        \If(\cbstart
	        \cbend){
	        $x^2_{t,i} \in \mathcal{C}$ and $b^l_t\le price(x_{t,i}) < b^u_t$ and $occ(x_{t,i})=oc$}
	        {
	            $I^p_t \leftarrow I^p_t \cup \{x_{t,i}\}$\Comment*[r]{\normalfont{update} $I^p_t$ with a new item}
	        }
	    }
    	$I^p_t \leftarrow$ retrieve $m_t$ nearest neighbors of the items in $I'_t$ from $I^p_t$\Comment*[r]{\normalfont{filter items from} $I^p_t$} 
		\textbf{return} $I^p_t$
\end{algorithm}

\subsection{Retrieval of Preferred Items}\label{ret_pref_items}
\cbstart
There are mainly three types of user preference modeling when providing recommendations to users, i.e., either content based or collaborative filtering based or combination of both. Since, collaborative filtering mostly suffers from the cold start problem in case of new users whose purchase history is not known, we employ a content based item retrieval method that retrieves relevant items, which are visually similar to user preferences.
Given a set of 
chosen items $I'_t$ and price-range $[b^l_t,  b^u_t)$ for clothing type $t$ corresponding to occasion $oc$, BOXREC retrieves a subset of items $I^p_t\subseteq I_t$ for each type $t\in \mathcal{T}$ such that the set $\mathcal{I}^p$ = $I^p_1 \cup I^p_2 \cup\ldots\cup I^p_{|\mathcal{T}|}$, where each $I^p_t$ is comprised of items suitable for the occasion $oc$ with price in the range $[b^l_t,b^u_t)$ and similar to items in $I'_t$ in terms of \textit{category} and \textit{visual} features.
\cbend

Algorithm \ref{alg:ret_items} illustrates the method for retrieving the set of preferred items of a particular item type $t$.
At first, $\mathcal{C}$ is assigned with the set of item categories (corresponding to type $t$) present in the items of $I'_t$ (Line 2) and $I^p_t$ is initialized to an empty set (Line 3). Next, for each item $x_{t,i}$ of $I_t$, the algorithm checks whether the category $x^2_{t,i}$ is in $\mathcal{C}$, $price(x_{t,i})$ is within the range $[b^l_t, b^u_t)$ and $occasion(x_{t,i})$ is the same as $oc$ (Line 5). If all the conditions at Line 5 are satisfied, then $x_{t,i}$ is included in $I^p_t$ (Line 6). Finally, a set of $m_t$ nearest neighbors of the items in $I'_t$ are filtered from $I^p_t$ (Line 9) and returned as output. 
\cbstart
\begin{figure}[t]
	\centering
	\includegraphics[trim=0 0 0 0,clip,width=\textwidth]{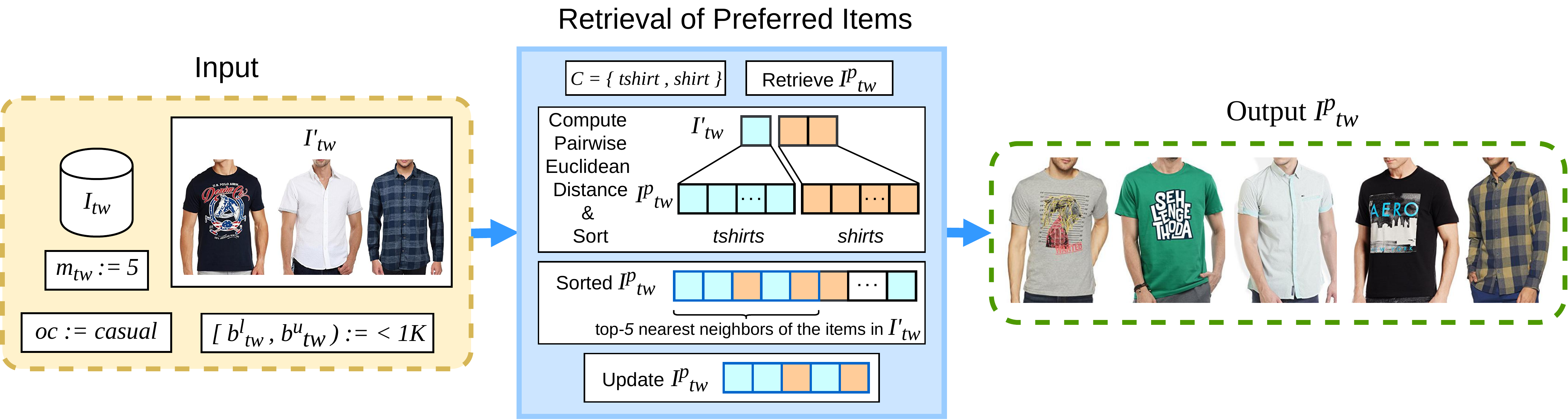}
	\caption{An overview for Retrieval of Preferred Top-wear Items}
	\label{fig:retrieval_of_pref_items}
\end{figure}
\cbend

\cbstart
As an example, Fig. \ref{fig:retrieval_of_pref_items} shows retrieval of preferred top-wear items, where the input $I_{tw}$ is the set of top-wear items present in the database, $I'_{tw}$ is the set of three chosen top-wears, occasion $oc$ is $casual$, price-range [$b^l_{tw},b^u_{tw}$) is $< 1$K and required number of top-wear items, i.e., $m_{tw}$ is set as $5$. 
 Now, $I'_{tw}$ has two clothing categories, i.e., $\mathcal{C}$ = \{$tshirt$, $shirt$\} and $I^p_{tw}$ is retrieved from $I_{tw}$ by satisfying category, price-range and occasion. Then we compute pairwise euclidean distance between the feature vectors of the items in $I_{tw}$ and $I'_{tw}$ that have matching category (e.g., only distance between pair of shirts or tshirts are computed and not between shirt and tshirt), followed by sorting (in ascending order) all the items in $I^p_{tw}$ based on the computed distances. The last step is updation of $I^p_{tw}$ with the top-$m_t$ (= $5$) top-wear items, which is the final output.

Algorithm \ref{alg:ret_items} uses a 2048-dimensional visual feature vector extracted from the linear layer output of GoogleNet Inception V3 architecture \cite{Szegedy:2016} (fine-tuned on fashion images)
corresponding to each item image $x^1_{t,i}$ for computing the distance between item pairs required for getting the nearest neighbors of the items in $I'_t$ from the set $I^p_t$. 
\cbend
\begin{figure}[h]
	\centering
	\includegraphics[trim=0 0 0 0,clip,width=0.7\textwidth]{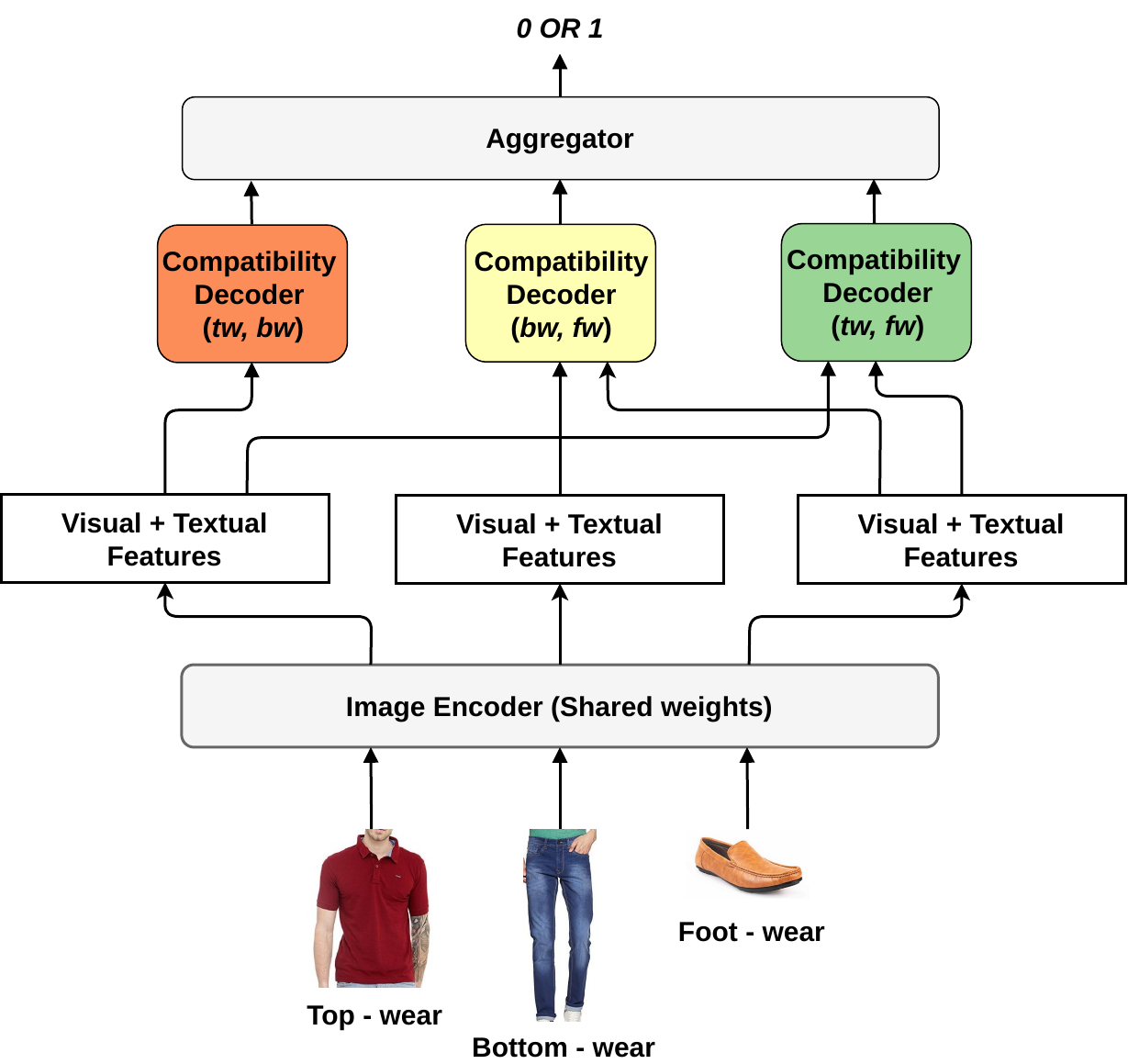}
	\caption{Overview of the proposed Outfit Scoring Framework}
	\label{fig:abstract_fig}
\end{figure}
\subsection{Outfit Scoring Framework}\label{outfit_scoring_framework}
The outfit scoring framework of BOXREC (BOXREC-OSF) verifies whether a set of items compose a good or a bad outfit. As shown in Fig. \ref{fig:abstract_fig}, BOXREC-OSF primarily consists of three components: a shared image encoder that provides a representation for items of different categories, separate compatibility decoders for type-specific item pairs and an aggregator that compiles the compatibility scores ($0$ or $1$) of all the item pairs in an outfit into a single binary $0$ or $1$ value. 

\cbstart
Outfit scoring is mainly focused on predicting pairwise compatibility between items based on visual and/or textual features. So, during the training phase, price of individual items is not required.
When modeling compatibility between a pair of items, most of the existing methods try to map the image features of the items into a latent space such that compatible items are close to each other in that space, while incompatible items are far apart. This is the same method used in modeling similar item retrieval systems. However, there is a clear distinction between compatibility and similarity. So, a different approach is required to model compatibility. 

Now, if we assign the task of judging the compatibility between a white graphic tshirt and a shaded blue denim jeans to a fashion expert. The fashion expert will easily point out that the graphic pattern in the tshirt gets highlighted when paired with the jeans, white and blue complements each other, etc. This human judgement is based on prediction of mutual compatibility between a pair of items (with respect to different aspects), and can be best approximated by incorporating mutual attention of visual features in the compatibility decoder. Our outfit scoring framework is an enhancement over the 
existing mutual attention based method proposed by Lin et al. \cite{Lin:2019} that lacks an expressive image encoder, and fails to make compatibility predictions when tested on items that are not present in the training dataset. We exploit the full potential of the mutual attention mechanism by introducing suitable modifications and train separate compatibility decoders for each item pair types. 
\cbend
The aggregator finally performs a logical AND operation on the binary matching scores of all the item pairs in an outfit to predict the final outfit score.  
\cbstart
\begin{figure}[t]
    \centering
	\includegraphics[trim=0 0 0 0,clip,width=0.9\textwidth]{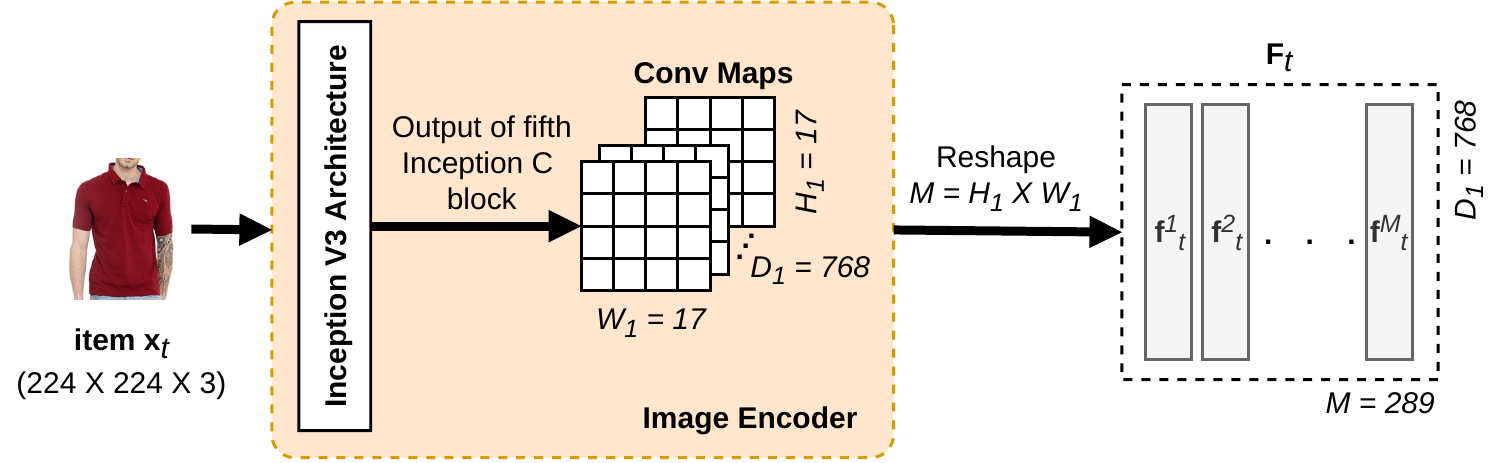}
	\caption{Architecture of Image Encoder
	}
	\label{fig:img_encoder}
\end{figure}
\cbend
\subsubsection{Image Encoder}
\cbstart
We encode each item image using the well known Convolutional Neural Network (CNN) architecture of Inception v3  \cite{Szegedy:2016}.
As hown in Fig. \ref{fig:img_encoder}, for an image $x^1_t$ of item $x_t$, we extract the convolutional feature map output of the fifth Inception C module\footnote{We used the PyTorch implementation: \url{https://pytorch.org/docs/stable/\_modules/torchvision/models/inception.html}}
fine-tuned on fashion images as $\bm{\mathrm{F}}_t \in \mathbb{R}^{D_1\times H_1\times W_1}$ 
. Next, we reshape $\bm{\mathrm{F}}_t$ by flattening the height and width as $M=H_1\times W_1$. For image $x^1_t$, we extract visual features as:
\begin{equation}\label{eqn:visual_features}
\bm{\mathrm{F}}_t = [\bm{\mathrm{f}}^1_t, \ldots, \bm{\mathrm{f}}^M_t],\hspace{0.25cm} \bm{\mathrm{f}}^j_t\in \mathbb{R}^{D_1}
\end{equation}
where $\bm{\mathrm{f}}^i_t$ is the visual feature vector corresponding to the location $i$ in the input image.
\cbend
\subsubsection{Compatibility Decoder and Aggregator of BOXREC-OSF}
\cbstart
Given a pair of images $x^1_a$ and $x^1_b$, 
where $a,b\in \mathcal{T}$, we first retrieve the visual features $\mathbf{F}_a$ = \{$\bm{\mathrm{f}}^1_a, \bm{\mathrm{f}}^2_a,\ldots,\bm{\mathrm{f}}^M_a$\} and $\mathbf{F}_b$ = \{$\bm{\mathrm{f}}^1_b, \bm{\mathrm{f}}^2_b,\ldots,\bm{\mathrm{f}}^M_b$\} following Equation \ref{eqn:visual_features}.
We then apply a global average pooling (GAP) to $\mathbf{F}_a$ and $\mathbf{F}_b$ for aggregating all the local visual features of the two images into global image embeddings $\bm{\mathrm{g}}_a \text{and}\  \bm{\mathrm{g}}_b \in \mathbb{R}^{D_1}$.
\begin{equation}\label{global_avg_pool}
    \bm{\mathrm{g}}_a = \frac{1}{M}\sum\limits_{j=1}^{M} \bm{\mathrm{f}}^j_a, \hspace{0.25cm}
    \bm{\mathrm{g}}_b = \frac{1}{M}\sum\limits_{j=1}^{M} \bm{\mathrm{f}}^j_b
\end{equation}
\cbstart
\begin{figure}[t]
	\centering
	\includegraphics[trim=0 0 0 0,clip,height=135mm]{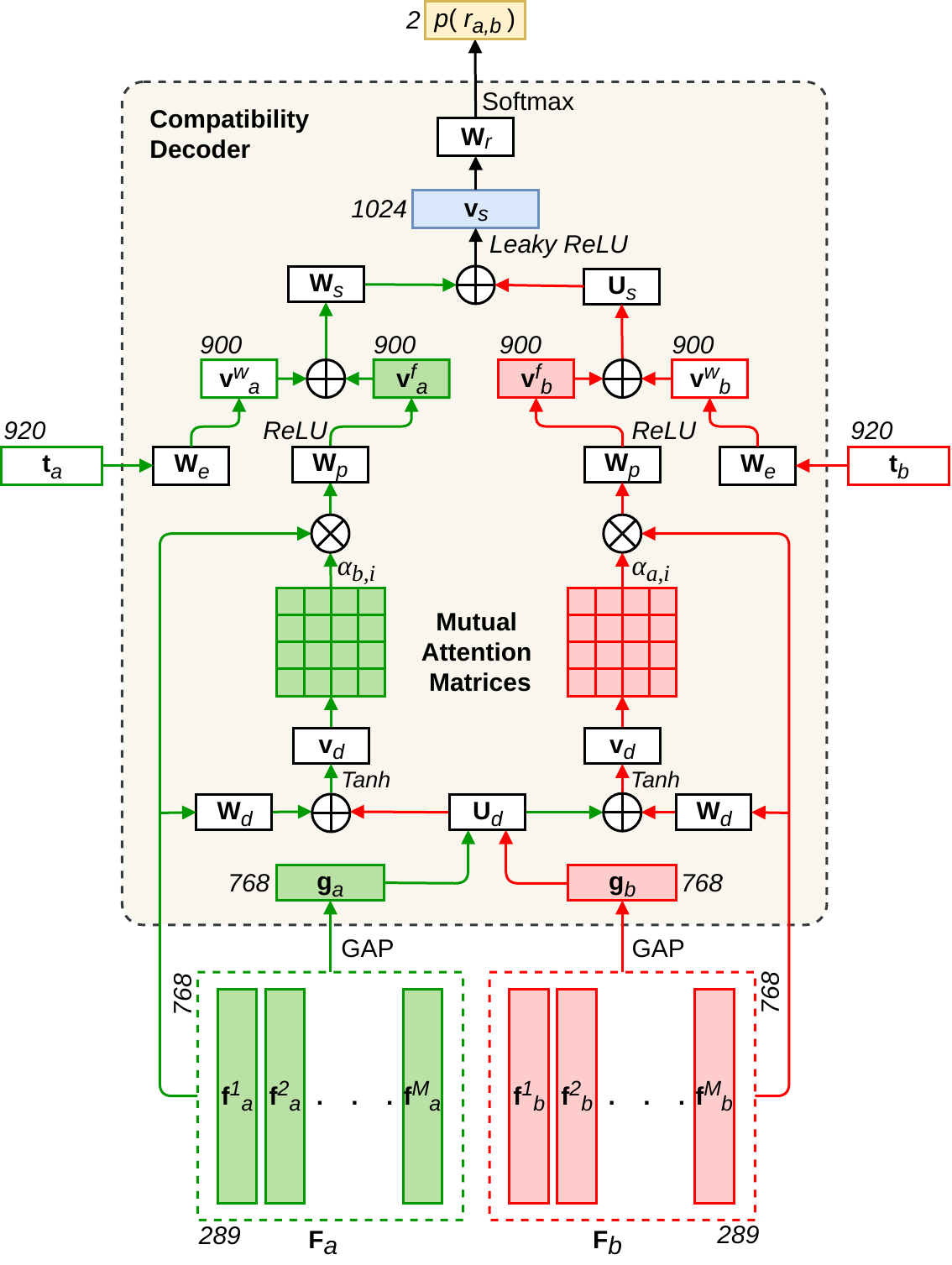}
	\caption{Architecture of Pairwise Compatibility Decoder}
	\label{fig:matching_decoder}
\end{figure}
\cbend
Next, we compute the
attention weight from $x_a$ to $x_b$ \cite{Bahdanau:2015} for the location $i$ of image $x_a$ as follows:
\cbend
\begin{equation}\label{eqn:attn_wts_cat_a}
\beta_{a,i} = \bm{\mathrm{v}}^\top_d \textit{tanh}(\mathbf{W}_d\bm{\mathrm{f}}^i_b + \
\mathbf{U}_d\bm{\mathrm{g}}_a)
\end{equation}
where $\mathbf{W}_d$ and $\mathbf{U}_d \in \mathbb{R}^{D_1\times D_1}$ and $\bm{\mathrm{v}}_d \in \mathbb{R}^{D_1}$. We then normalize the attention weights using Equation \ref{eqn:norm_attn_wts_cat_a}.
\begin{equation}\label{eqn:norm_attn_wts_cat_a}
\alpha_{a,i} = \frac{\text{exp}(\beta_{a,i})}{\sum_{i=1}^M \text{exp}(\beta_{a,i})}
\end{equation}
We get the attentive visual features $\hat{\bm{\mathrm{g}}}_b$ of $x_b$ as:
\begin{equation}
\hat{\bm{\mathrm{g}}}_b = \sum_{i=1}^M \alpha_{a,i}\bm{\mathrm{f}}^i_b
\end{equation}
Similarly, we compute: 
\begin{subequations}
\begin{align}
\beta_{b,i} &= \bm{\mathrm{v}}^\top_d \text{tanh}(\mathbf{W}_d\bm{\mathrm{f}}^i_a + \
\mathbf{U}_d\bm{\mathrm{g}}_b)
\\
\alpha_{b,i} &= \frac{\text{exp}(\beta_{b,i})}{\sum_{i=1}^M \text{exp}(\beta_{b,i})}
\\
\hat{\bm{\mathrm{g}}}_a &= \sum_{i=1}^M \alpha_{b,i}\bm{\mathrm{f}}^i_a
\end{align}
\end{subequations}
Finally, we project $\hat{\bm{\mathrm{g}}}_a$ and $\hat{\bm{\mathrm{g}}}_b$ to visual feature vectors $\bm{\mathrm{v}}^f_a$ and $\bm{\mathrm{v}}^f_b \in \mathbb{R}^{A_1}$:
\begin{equation}
\bm{\mathrm{v}}^f_a = \text{ReLU}(\mathbf{W}_p \hat{\bm{\mathrm{g}}}_a), \hspace{0.25cm}
\bm{\mathrm{v}}^f_b = \text{ReLU}(\mathbf{W}_p \hat{\bm{\mathrm{g}}}_b)
\end{equation}
where $\mathbf{W}_p \in \mathbb{R}^{A_1\times D_1}$.
\cbstart
Apart from this, we also use textual metadata, i.e., category, product title/description as bag of words. Let, $V$ be the vocabulary of all words that are present in the textual metadata  of all fashion items in the dataset. We create a binary vector $\mathbf{t}_a\in \{0,1\}^{|V|}$ to represent the descriptive words associated with $x_a$. Similarly, we create $\mathbf{t}_b\in \{0,1\}^{|V|}$ corresponding to item $x_b$. Both are then transformed to textual feature vectors $\mathbf{v}^w_a$ and $\mathbf{v}^w_b\in \mathbb{R}^{A_1}$:
\begin{equation}
    \mathbf{v}^w_a = \mathbf{W}_e \mathbf{t}_a,\ \mathbf{v}^w_b = \mathbf{W}_e \mathbf{t}_b
\end{equation}
where transformation matrix $\mathbf{W}_e\in \mathbb{R}^{A_1\times|V|}$.
Now, there are different ways of combining neural network features \cite{KedanLi:2019}. We preform summation of visual and textual feature vectors to get the latent representations $\bm{\mathrm{v}}_a$ and $\bm{\mathrm{v}}_b \in \mathbb{R}^{A_1}$:
\begin{equation}\label{latent_repr}
\bm{\mathrm{v}}_a = tanh(\bm{\mathrm{v}}^f_a + \bm{\mathrm{v}}^w_a),\hspace{0.25cm} \bm{\mathrm{v}}_b = tanh(\bm{\mathrm{v}}^f_b + \bm{\mathrm{v}}^w_b)
\end{equation}


A multi-layer neural network is used to determine the matching probability of $x_a$ and $x_b$.
Latent representations $\bm{\mathrm{v}}_a$ and 
$\bm{\mathrm{v}}_b$ are first mapped into a shared space $\bm{\mathrm{v}}_s$. 
\begin{equation}
\bm{\mathrm{v}}_s = \textit{LeakyReLU}(\mathbf{W}_s\bm{\mathrm{v}}_a + \mathbf{U}_s\bm{\mathrm{v}}_b)
\end{equation}
where $\bm{\mathrm{v}}_s \in \mathbb{R}^{B_1}$, $\mathbf{W}_s$ and $\mathbf{U}_s \in \mathbb{R}^{{B_1}\times A_1}$.
Finally, the compatibility probability is estimated as follows:
\begin{equation}
p(r_{a,b}) = softmax(\mathbf{W}_r \bm{\mathrm{v}}_s)
\end{equation}
where $\mathbf{W}_r \in \mathbb{R}^{2\times B_1}$ and $p(r_{a,b}) \in \mathbb{R}^2$ which provide two probabilities for $r_{a,b}=0$ ($x_a$ and $x_b$ do not match) and $r_{a,b}=1$ ($x_a$ and $x_b$ are compatible).
\cbend
If $p(r_{a,b}=1)>p(r_{a,b}=0)$, $score(x_a,x_b;\theta^{(a,b)})=1$; otherwise, $score(x_a,x_b;\theta^{(a,b)})=0$. As mentioned in \cite{Vasileva:2018:LTAE}, respecting types is important in computing the compatibility of a pair of items. So we train separate compatibility decoders for each type pairs, e.g., ($tw, bw$), ($bw, fw$) and ($tw, fw$).  

Now, to predict whether a particular set of items constitute an outfit, we logically AND the pairwise compatibility scores of all the item pairs as follows:
\begin{equation}
    \psi(o_j;\theta) = \bigwedge_{x_a,x_b\in o_j}score(x_a,x_b;\theta^{(a,b)})
\end{equation}
where $\theta^{(a,b)}$ denotes the model parameters for pair type $(a,b)$ and $\theta = \cup_{a,b\in \mathcal{T}} \theta^{(a,b)}$.
\subsubsection{Learning Framework of BOXREC-OSF}
\cbstart
We use negative log-likelihood loss function for
the compatibility decoders as follows:
\begin{equation}
    L^{(a,b)}_{compat} = \sum_{r_{a,b}:(x_a,x_b)\in \mathcal{D}^{+}_{a,b}\cup \mathcal{D}^{-}_{a,b}}^N -log(p(r_{a,b}))
\end{equation}
where $\mathcal{D}^{+}_{a,b}$ and $\mathcal{D}^{-}_{a,b}$ are the set of positive and negative item pairs corresponding to item types $a$ and $b$, respectively. 
\cbend
In order to avoid overfitting, we also include L2 loss regularization:
\begin{equation}
    L^{(a,b)}_{reg} = \Vert\theta^{(a,b)}\Vert^2_2
\end{equation}
\cbstart
We also compute visual semantic embedding loss to project visual and textual features close to each other as follows:
\begin{equation}
    L^{(a,b)}_{vse} = \Vert\mathbf{v}^f_a - \mathbf{v}^w_a\Vert^2_2 + \Vert\mathbf{v}^f_b - \mathbf{v}^w_b\Vert^2_2
\end{equation}
Finally, the objective function for compatibility decoder of the pair type ($a,b$) is a linear combination of $L^{(a,b)}_{compat}$, $L^{(a,b)}_{reg}$ and $L^{(a,b)}_{vse}$:
\begin{equation}
    L = L^{(a,b)}_{compat} + \lambda_1 L^{(a,b)}_{reg} + \lambda_2 L^{(a,b)}_{vse}
\end{equation}
\cbend
where $\lambda_1$ and $\lambda_2$ are scalar hyperparameters.

\cbstart
After collecting the user preferences across different item types and generating the set of preferred outfits $O^p$, the final step of BOXREC is to recommend a box full of items that compose maximum possible preferred outfit combinations, satisfying the overall shopping budget. 
\cbend

\subsection{Recommendation of Items and Outfits}\label{reco_step}
\cbstart
For a particular shopping session, a user mentions a maximum total budget for shopping a set of items that compose different good looking outfit combinations. Since $O^p$ covers a wide range of items, the cost of all the items in $O^p$ may exceed user's budget. So, BOXREC recommends a subset $H\subseteq O^p$, such that the total cost $T(H)$ is at most $B$ and the number of outfits in $H$ is maximum\footnote{$O^p$ is a set of sets, i.e., each element is a set of items that form an outfit. While maximizing the size of set $H$, outfits $o\in O^p$ need to be added to $H$ one at time, such that $T(H)\le B$.}.
\cbend
This is an optimization problem~-~Box Recommendation with Maximum Outfits that we formally define as follows. 
\begin{defn}\textit{Box Recommendation with Maximum Outfits (BOXRECMO) problem~-~}Given a finite set of items $I$ = \{$x_1, x_2,\ldots,x_N$\}, a finite collection of non-empty preferred outfit sets $O^p$ = \{$o_1, o_2,\ldots,o_L$\}, where each set $o_j$ is an outfit having a subset of items from $I$, and a positive integer $B$, indicating the maximum total shopping budget, find a subset $H \subseteq O^p$, such that the total price of the distinct items in $H$ is at most $B$ and $|H|$ is maximum, i.e., to
\begin{equation*}
maximize\hspace{0.1cm} |H|
\end{equation*}
\begin{equation*}
s.t. \hspace{0.1cm}T(H) \le B
\end{equation*}
\end{defn}
\noindent
Here, we assume that $I = \nu(O^p)$, $price(o_j) \le B$ and $|o_j| = |\mathcal{T}|$, $\forall o_j \in O^p$. Thus, maximizing $|H|$ is equivalent to maximizing $|\mathcal{T}|\times|H|$, or simply $Card(H)$. 
Thus, we can rewrite the objective of BOXRECMO as follows:
\begin{equation*}
maximize\hspace{0.1cm} Card(H)
\end{equation*}
\begin{equation*}
s.t. \hspace{0.1cm}T(H) \le B
\end{equation*}
We prove that the BOXRECMO problem is NP-Complete by showing a decision version of BOXRECMO to be in NP and proving that it is NP-Hard by reducing a decision version of Fusion Knapsack problem \cite{Grange:2015} in polynomial time to the decision version of BOXRECMO problem.
\begin{defn}\textit{Decision version of BOXRECMO (D-BOXRECMO) problem~-~}Given a finite set of items $I$ = \{$x_1, x_2,\ldots,x_N$\}, a finite collection of non-empty preferred outfit sets $O^p$ = \{$o_1$, $o_2$, $\ldots,o_L$\}, where each set $o_j$ is an outfit having a subset of items from $I$, a positive integer $B$ indicating maximum total shopping budget and a positive integer $W$, is there any subset $H \subseteq O^p$ such that 
\begin{equation}\label{eqn:d-boxrec-constraint}
T(H) \le B
\end{equation}
and
\begin{equation}\label{eqn:d-boxrec-maxim}
Card(H)\ge W
\end{equation}
\end{defn}
\begin{defn}\textit{Fusion Knapsack (FKS) problem~-~}Given a finite set of symbols $\mathcal{A}$ = \{$a_1$, $a_2$, $\ldots$ , $a_m$\}, a finite collection of nonempty sets $\mathcal{S}$ = \{$s_1, s_2, \ldots, s_n$\}, where each set $s_i$ is a tile having a subset of symbols from $\mathcal{A}$ and positive integer capacity $C$, find a subset $S$ of $\mathcal{S}$ so that the 
number of distinct symbols present in the union of all the tiles in $S$ is at most $C$ and sum of the size of all the tiles in $S$ is maximum, i.e., to 
\begin{equation*}
    maximize \sum\limits_{s_i \in S}|s_i|
\end{equation*}
\begin{equation*}
s.t. \biggl|\bigcup\limits_{s_i\in S}{s_i}\biggr| \le C
\end{equation*}

\end{defn}
\begin{defn}\textit{Decision version of Fusion Knapsack (D-FKS) problem~-~}Given a finite set of symbols $\mathcal{A}$ = \{$a_1, a_2, \ldots, a_m$\}, a finite collection of nonempty sets $\mathcal{S}$ = \{$s_1, s_2, \ldots, s_n$\}, where each set $s_i$ is a tile having a subset of symbols from $\mathcal{A}$, a positive integer capacity $C$ and a positive integer $Y$, is there a subset $S \subseteq \mathcal{S}$ such that
\begin{equation*}
    \biggl|\bigcup\limits_{s_i\in S}{s_i}\biggr| \le C
\end{equation*}and
\begin{equation}
    \sum\limits_{s_i \in S}|s_i| \ge Y
\end{equation}
\end{defn}
\noindent
\begin{thm}
D-BOXRECMO is NP-Complete.
\end{thm}
\noindent
\begin{proof}
Given a subset $H \subseteq O^p$, positive integers $B$ and $W$, condition (\ref{eqn:d-boxrec-constraint}) can be verified in polynomial time by computing the total price $T(H)$ 
and comparing it with $B$. We can also verify condition (\ref{eqn:d-boxrec-maxim}) in polynomial time by computing the cardinality $Card(H)$ (i.e., the number of items distinct or not in $H$) and comparing it with $W$. Hence, D-BOXRECMO is in NP.

Now, we prove that D-FSK $\le_p$ D-BOXRECMO, i.e., there is a polynomial time reduction from any instance of D-FSK to some instance of D-BOXRECMO. It suffices to show that there exists a polynomial time reduction $\mathcal{Q}(\cdot)$, such that, $\mathcal{Q}(X)$ is 'Yes' instance to the D-BOXRECMO problem iff $X$ is a 'Yes' to the D-FSK problem. Suppose we are given a set $\mathcal{A}$ = \{$a_1, a_2, \ldots, a_m$\}, $\mathcal{S}$ = \{ $s_1, s_2, \ldots, s_n$\}, capacity $C$ and integer $Y$ for any instance $X$ of the D-FSK problem. Consider the following instance $\mathcal{Q}(X)$ of the D-BOXRECMO problem: $x_i = a_i$, $o_j= s_j$, $B = C$, $W = Y$, $L=n$ and $price(x_i)=1\ \text{for each item}\ x_i \in o_j$ ($\forall o_j \in O^p$). 
Firstly, if the answer for $X$ is 'Yes' for the D-FKS problem, i.e., there exists a subset $S\subseteq \mathcal{S}$, such that $|\cup_{s_j\in S}s_j| \le C$ and $\sum_{s_j\in S} |s_j| \ge Y$, then for $\mathcal{Q}(X)$, we take $H = S$. Thus, we have:
\begin{enumerate}
    \item $T(H) = \sum_{x_i\in \nu(H)}price(x_i) = |\cup_{o_j\in H}o_j| = |\cup_{s_j\in S}s_j| \le C = B$,
    \item $Card(H) = \sum_{o_j\in H}|o_j| = \sum_{s_j\in S}|s_j|\ge Y = W$.
\end{enumerate}
This shows that the answer for $\mathcal{Q}(X)$ is also 'Yes'.

Conversely, suppose that the answer for $\mathcal{Q}(X)$ is 'Yes' for the D-BOXRECMO problem, that is, there exists a subset $H\subseteq O^p$, satisfying conditions (\ref{eqn:d-boxrec-constraint}) and (\ref{eqn:d-boxrec-maxim}). For $X$, we take $S=H$. From (\ref{eqn:d-boxrec-constraint}) and (\ref{eqn:d-boxrec-maxim}) we have $|\cup_{s_j\in S}s_j| = |\cup_{o_j\in H} o_j| = \sum_{x_i\in \nu(H)}price(x_i) = T(H) \le B = C$ and $\sum_{s_j\in S}|s_j| = \sum_{o_j\in H}|o_j| = Card(H) \ge W = Y$. Thus, $X$ is the 'Yes' instance of the D-FSK problem. 
Hence, D-BOXRECMO is NP-Hard. 

Since D-BOXRECMO is both in NP and is NP-Hard, it is NP-Complete.
\end{proof}

\section{Heuristic Approach to solve BOXRECMO}\label{heuristics}
Since BOXRECMO is an NP-Complete problem, 
a heuristic approach is required to solve it.
We suitably adopt the Overload and Remove (OLR) heuristic
\cite{Grange:2018} in our proposed solution for BOXRECMO. The OLR heuristic is a non-greedy approach to solve the NP-Complete problem of PAGINATION \cite{Grange:2018}, which is a generalization of the well known BIN PACKING problem \cite{Martello:1990:BinPacking}.
The solution generated by OLR for the PAGINATION problem is further refined by a decantation process, which attempts to reduce the number of pages.

Decantation involves three steps, which settle at the beginning of the pagination as many pages, connected components and tiles as possible. Two tiles are said to be connected if and only if they share a common symbol between them (e.g., tiles $\{a_1, a_2, a_3\}$ and $\{a_3, a_4, a_5\}$ are connected by the symbol $a_3$ ) or are both connected to an intermediate tile (e.g. tiles $\{a_1, a_2, a_3, a_4, a_5\}$  and $\{a_5, a_6, a_7\}$ are connected by the intermediate tile  $\{a_4, a_5, a_6\}$).
After decantation of a valid pagination on the pages (components/tiles), no page (component/tile) contents can be shifted to a page (component/tile) having lesser index without making the pagination invalid. 

The problem of PAGINATION can be formally defined as follows:
\begin{defn}\textit{PAGINATION~-~}
Given a finite set of symbols $A$ = \{$a_1, a_2,\ldots,a_m$\}, a finite collection of non-empty finite set of tiles $\mathcal{S}$ = \{$s_1, s_2,\ldots,s_n$\}, where each set $s_j$ is a tile having a subset of symbols from $A$, and a positive integer $C$, indicating the capacity of each page $p$, find a partition (or pagination) $\mathcal{P}$ of $\mathcal{S}$ such that for any page $p\in \mathcal{P}$ total number of tiles in $p$ is at most $C$ and number of pages in $\mathcal{P}$ is minimum, i.e., to 
\begin{equation*}
    minimize |\mathcal{P}|
\end{equation*}
\begin{equation*}
    s.t. \biggl| \bigcup\limits_{t\in p}{s}\biggr | \le C, \hspace{0.1cm}\forall p\in \mathcal{P}
\end{equation*}
\end{defn}
\noindent
where $|\mathcal{P}|$ denotes the number of pages in $\mathcal{P}$. Minimizing $|\mathcal{P}|$ is equivalent to packing as many overlapped tiles in each page $p$ as possible. In order to use the heuristic solution of PAGINATION to solve BOXRECMO, we show that any instance of BOXRECMO can be mapped to some instance of PAGINATION. 

Let, $<I, O^p, B, H>$ be an instance of BOXRECMO, where $I$ = \{$x_1, x_2,\ldots,x_N$\} is the set of items, $O^p$ = \{$o_1, o_2,\ldots,o_L$\} is the set of preferred outfits with each $o_j$ having a subset of items from $I$, $B$ is the total shopping budget and $H\in O^p$ is the box (or set) with maximum number of outfits, satisfying the budget constraint $B$. 

Now, consider the instance $<A, \mathcal{S}, C, \mathcal{P}>$ of PAGINATION, where $A$ = \{$a_1, a_2,\ldots,a_m$\} is the set of symbols with each $a_i = x_i$, $\mathcal{S}$ = \{$s_1, s_2,\ldots,s_n$\} is the set of tiles with each tile $s_j = o_j$, capacity of each page $p$ is $C=B$ and $\mathcal{P}$ is the partition (or pagination) of $\mathcal{S}$ with minimum number of pages, such that $P' = H$ 
(where $P'\in \mathcal{P}$ is the page with maximum number of tiles). Hence, an algorithm that solves PAGINATION can be used to compute the solution for BOXRECMO.


\begin{exmp}
Let $\mathcal{P}_1 = \{\{s_1\}, \{s_2\}, \{s_3, s_4\}\}$ be a valid pagination for the finite collection of non-empty set of tiles $\mathcal{S}$ = \{$s_1, s_2, s_3, s_4$\}, where tiles $s_1$ = $\{a_1, a_2, a_3\}$, $s_2$ = $\{a_2, a_3, a_4\}$, $s_3$ = $\{a_3, a_4, a_5\}$, $s_4$ = $\{a_6, a_7, a_8, a_9\}$ and a capacity $C = 5$. Pagination $\mathcal{P}_1$ can be refined to $\mathcal{P}_2$ = \{$\{s_1, s_2\}, \{s_3, s_4\}$\} by the decantation of $\mathcal{P}_1$ on the pages. Now, $\mathcal{P}_2$ is decanted on both pages and connected components. Finally, decantation of $\mathcal{P}_2$ on the tiles produces the pagination $\mathcal{P}_3$ = \{$\{s_1, s_2, s_3\}, \{s_4\}$\}, which is a fully decanted pagination. $\mathcal{P}_3$ cannot be further decanted on any of pages, connected components or tiles. 
\end{exmp}
\begin{algorithm}[ht]
	\caption{Overload and Remove with Decantation}
	\label{alg:overload_and_remove}
	\SetAlgoNoLine
	\SetArgSty{textnormal}
	\DontPrintSemicolon
	\KwIn{A finite collection of non-empty preferred outfit sets $O^p$ = \{$o_1, o_2,\ldots,o_n$\}. Maximum total shopping budget $B$.} 
	\KwOut{Box $H\subseteq O^p$.}
	\Begin{
		$Q \leftarrow$ queue containing all the outfits of $O^p$\\
		$\mathcal{H} \leftarrow \{\{\ \}\}$\\
		\While(\Comment*[f]{\normalfont{iterate until $Q$ becomes empty}}){$Q$ is nonempty}{
			$o \leftarrow dequeue(Q)$\Comment*[r]{\normalfont{obtain an outfit from $Q$}}
			$\mathcal{H}_o \leftarrow$ boxes of $\mathcal{H}$ where $o$ has never been put on\Comment*[r]{\normalfont{obtain new boxes for $o$}}
			\If(\Comment*[f]{\normalfont{check if no box in $\mathcal{H}_o$ to insert $o$}}){$\mathcal{H}_o$ \text{has no box} $H'$ \text{such that} $|o|_{H'}<|o|$}{
			 $\mathcal{H} \leftarrow \mathcal{H} \cup \{\{o\}\}$	\Comment*[r]{\normalfont{add to $\mathcal{H}$ a new box containing only $o$}}
				\textbf{continue with next iteration}\\
			}
			$H' \leftarrow$ box $H'$ of $\mathcal{H}_o$ such that $|o|_{H'}$ is minimal\Comment*[r]{\normalfont{obtain a box from $\mathcal{H}_o$ to insert $o$}}
		    $H' \leftarrow H' \cup \{o\}$\Comment*[r]{\normalfont{put $o$ in $H'$}}
			\While(\Comment*[f]{\normalfont{iterate until budget is satisfied}}){$T(H') > B$}{
				\textbf{remove} from $H'$ one outfit $o'$ minimizing $\frac{|o'|}{|o'|_{H'}}$\Comment*[r]{\normalfont{selectively remove outfits}}
				$enqueue(Q,o')$\Comment*[r]{\normalfont{put removed outfit in $Q$}}
			}  
		}
		\textbf{decantate} $\mathcal{H}$ on boxes, connected components and outfits, respectively\Comment*[r]{\normalfont{refine $\mathcal{H}$}} 
		$H \leftarrow$ box $H$ of $\mathcal{H}$, such that $|H|$ is maximum\Comment*[r]{\normalfont{obtain box with maximum outfits}}
		\textbf{return} $H$
	}	
\end{algorithm}
\noindent
Algorithm \ref{alg:overload_and_remove} appropriately adopts the OLR heuristic and applies the decantation process to solve BOXRECMO in two steps, as enumerated below:
\\\\
\noindent
\textbf{Step 1. Overload and Remove (OLR) Heuristic}\\
Given a finite collection of non-empty preferred outfit sets $O^p$ = \{$o_1, o_2,\ldots,o_n$\} and the maximum total shopping budget $B$, Algorithm \ref{alg:overload_and_remove}
returns a box $H$ (where $H \subseteq O^p$), such that $H$ contains maximum number of outfits and total price of distinct items in $H$, i.e., $T(H) \le B$.

At first, $Q$ (FIFO queue) and $\mathcal{H}$ are initialized with all the outfits in $O^p$ (Line 2) and a set having an empty box\footnote{Here a box corresponds to a set.} (Line 3), respectively. For every outfit $o$ fetched from $Q$ (Line 5), $\mathcal{H}_o$ keeps track of the boxes in $\mathcal{H}$ where $o$ has never been inserted. 
If there is no box $H'$ in $\mathcal{H}_o$ such that the relative size of outfit $o$ in $H'$ is less than the size of $o$ (Line 7), a new box consisting solely of $o$ is added to $\mathcal{H}$ (Line 8) and further execution is continued (Line 9) with the next iteration of the main loop (Line 4). 
If the condition at Line 7 is not satisfied, a box $H'\in \mathcal{H}$ is obtained (Line 11) for insertion of outfit $o$ (Line 12), such that $o$ has the minimal relative size in $H'$. 
If total price $T(H')$ exceeds $B$ on insertion of an outfit in $H'$, then an outfit $o'$ of strictly smallest $\frac{size}{relative\ size}$ ratio is removed from $H'$ (Line 14) and added to $Q$ (Line 15). To ensure termination, the removed outfit $o'$ is forbidden to re-enter the same box. If there are more than one outfits with the same $\frac{size}{relative\ size}$ ratio, we remove any one of them. The inner loop (Lines 13 to 16) terminates when $T(H') \le B$ and the main loop (Lines 4 to 17) terminates when $Q$ becomes empty. 
\\\\
\textbf{Step 2. Refinement by Decantation}\\
After completion of Step 1, $\mathcal{H}$ consists of a set of boxes, each filled with outfits taken from $O^p$. Next, $\mathcal{H}$ is refined by decanting on boxes, connected components and outfits, respectively (Line 18).  Finally, Algorithm \ref{alg:overload_and_remove} obtains the box $H \in \mathcal{H}$ having maximum number of outfits (Line 19) and terminates by returning $H$ (Line 20).

In the worst case, a particular outfit may overload all the boxes and get removed from each. So, $|O^p|$ is the maximum number of times an outfit may get removed. Now, one set intersection is required to test whether an outfit can be put in a box or not. 
\cbstart
Hence, the overall time complexity is $\mathcal{O}(|O^p|^3Card(O^p))$.
\cbend

\section{Experimental Evaluation}\label{exp_eval}
This section elaborates on data collection, implementation details of item retrieval, generation of outfits, outfit scoring framework~-BOXREC-OSF and comparative evaluation of BOXREC-OSF with respect to existing state of the art baselines along with an ablation test analysis as well as evaluation and analysis of recommendations provided by BOXREC.
\subsection{Datasets and Annotation}
\cbstart
We use the Product Advertising API of Amazon and collect three types of male clothing catalog items: top-wear, bottom-wear and foot-wear (for casual and formal occasions) from the Amazon India website\footnote{\url{www.amazon.in}}. 
There is an available 
Fashion Product Images Dataset\footnote{\url{https://www.kaggle.com/paramaggarwal/fashion-product-images-dataset}} that contains $44$K fashion catalog items (male, female and kids) of Myntra\footnote{\url{https://www.myntra.com/}} covering different clothing types, occasions and seasons. We consider casual and formal male top-wear, bottom-wear and foot-wear items from this dataset. 
As shown in Table \ref{table:dataset_stats}, there are $1457+3836 = 5293$ top-wear, $227+956 = 1183$ bottom-wear and $1250 + 2539 =3789$ foot-wear items of (Amazon + Myntra), where each item is tagged with type, category and price (in Indian currency).

Since the (Amazon + Myntra) item dataset is not sufficient enough to train a classifier for the clothing category recognition task, we collect some more data across different male clothing types and categories from  Zalando\footnote{\url{www.zalando.co.uk}}, a famous fashion e-commerce site.
The combined (Amazon + Myntra + Zalando) item data are split into the ratio of $80\%$ (train) and $20\%$ (validation) for fine-tuning the architecture of the image encoder on fashion images covering 27 item categories across different item types. 

\begin{figure}[t]
	\centering     
	\subfigure[]{\label{subfig:annotation_desktop}\includegraphics[height=75mm]{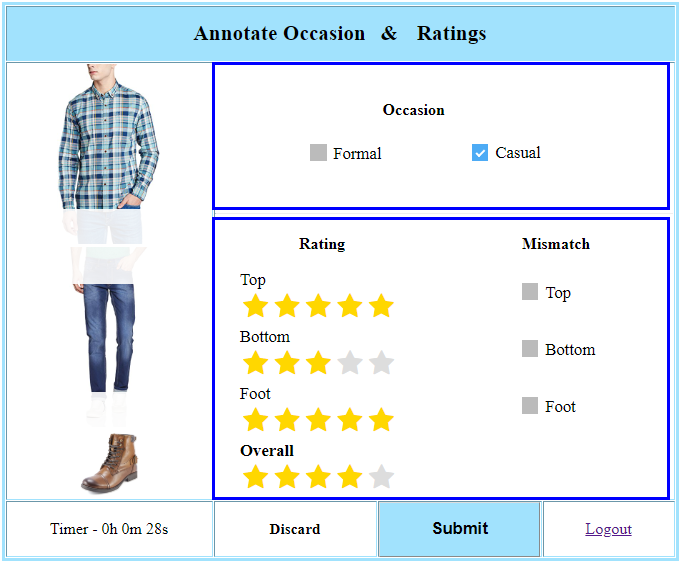}}
	\subfigure[]{\label{subfig:annotation_mobile}\includegraphics[height=75mm]{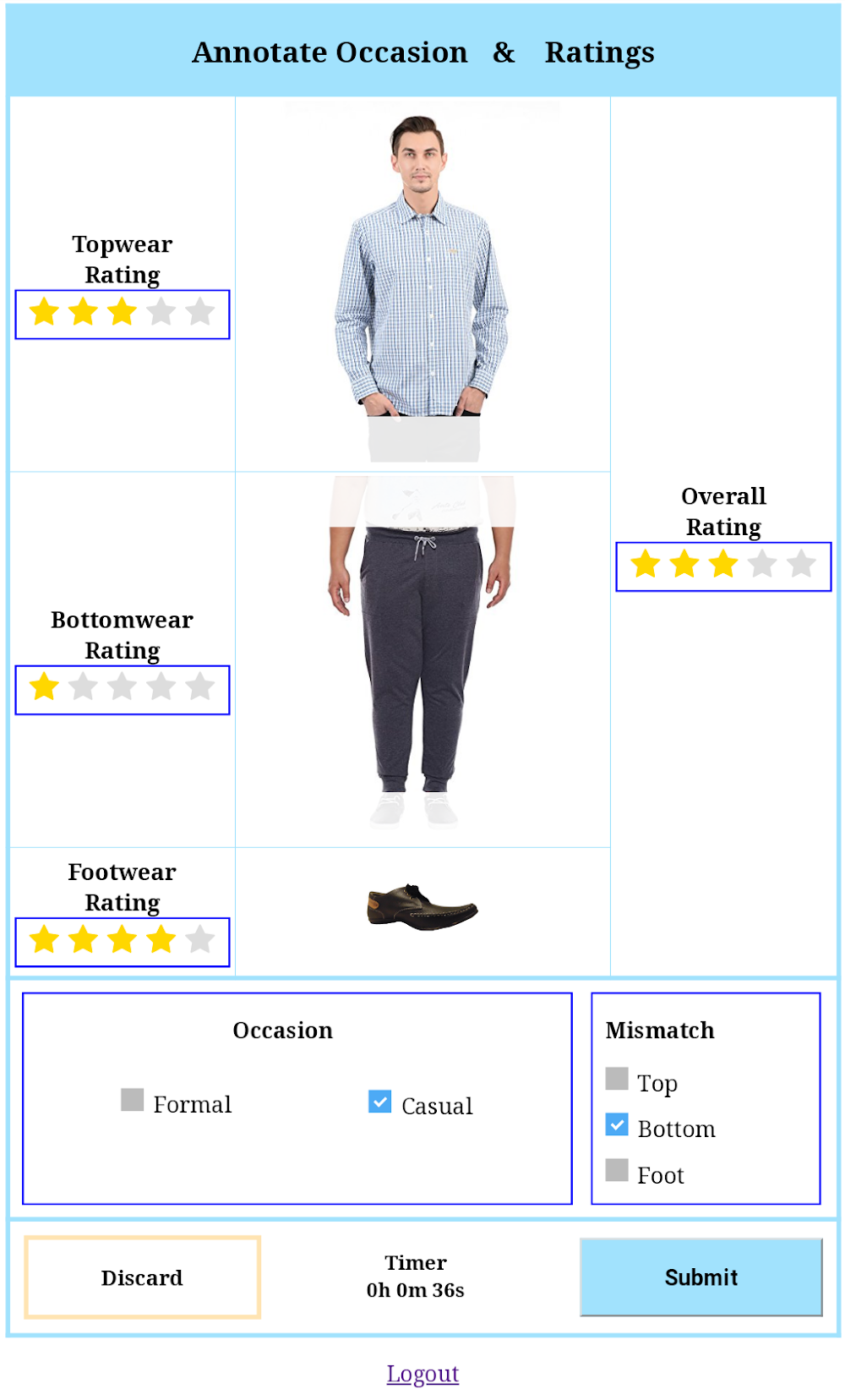}}
	\caption{Annotation Interface: a) Desktop version, b) Mobile device version}\label{fig:annotation}
\end{figure}

We create $30000$ and $15000$ distinct combinations by randomly combining fashion items from the item dataset of Amazon and Myntra, respectively. Each combination consists of three items (one from each of top-wear, bottom-wear and foot-wear item types). We  
manually annotate them using our own online interface as shown in Fig. \ref{fig:annotation}.
But some of the combinations are discarded because of the presence of some discrepancies
(such as missing item, presence of female clothing item, bottom-wear in place of top-wear, etc.). So, we refine both (Amazon + Myntra) item and outfit dataset by discarding invalid items and outfits, respectively. The details of items, item pairs and outfits that are shown in Table \ref{table:dataset_stats} present the updated dataset information. After updation, we have $27763+14769 = 42532$ (Amazon + Myntra) combinations.

\begin {table*}[t]
\caption {Details of E-commerce Datasets}\label{table:dataset_stats}
\centering
\begin{tabular}{c | c | c c c}\toprule
\multirow{2}{*}{\textbf{type}} & \multirow{2}{*}{\textbf{category}} &
\multicolumn{3}{c}{\textbf{No. of Items}}\\
\cline{3-5}
& & Amazon & Myntra & Zalando \\
\hline
\multirow{5}{*}{\textit{top-wear}}
& \textit{shirt} & 761 & 1811 & 1546\\
\cline{2-5}
& \textit{tshirt} & 295 & 1869 & 3915\\
\cline{2-5} 
& \textit{polo tshirt} & 311 & 85 & 1071\\
\cline{2-5}
& \textit{long sleeved top} & 90 & 71 & 732\\
\cline{2-5}
& \textbf{Total} & \textbf{1457} & \textbf{3836} & \textbf{7264}\\
\hline
\multirow{5}{*}{\textit{outer-wear}} 
& \textit{blazer} & - & - & 240\\
\cline{2-5} 
& \textit{coat} & - & - & 281 \\
\cline{2-5}
& \textit{jacket} & - & - & 2210 \\
\cline{2-5}
& \textit{jumper/cardigan} & - & -& 3383\\
\cline{2-5}
& \textbf{Total} & - & - & \textbf{6114} \\
\hline
\multirow{5}{*}{\textit{bottom-wear}} 
& \textit{trouser/chino} & 139 & 409 & 1182\\
\cline{2-5} 
& \textit{jeans} & 5 & 297 & 2409 \\
\cline{2-5}
& \textit{track-pant} & 77 & 80 & 756\\
\cline{2-5}
& \textit{shorts} & 6 & 170 & 1019\\
\cline{2-5}
& \textbf{Total} & \textbf{227} & \textbf{956} & \textbf{5366}\\
\hline
\textit{top-bottom-combo} & suit & - & - & 203 \\
\hline
\multirow{5}{*}{\textit{foot-wear}} 
& \textit{ankle-boot} & 177 & 76 & 162\\
\cline{2-5} 
& \textit{lace-up} & 145 & 195 & 209 \\
\cline{2-5}
& \textit{slip-on} & 554 & 816 & 56 \\
\cline{2-5}
& \textit{trainer} & 374 & 429  & 751\\
\cline{2-5}
& \textit{sandals} & - & 1023 & -\\
\cline{2-5}
& \textbf{Total} & \textbf{1250} & \textbf{2539} & \textbf{1178}\\
\hline
\multirow{10}{*}{\textit{accessory}} 
& \textit{backpack} & - & - & 390\\
\cline{2-5} 
& \textit{bag} & - & - & 932 \\
\cline{2-5}
& \textit{belt} & - & - & 293 \\
\cline{2-5}
& \textit{head-wear} & - & - & 191\\
\cline{2-5}
& \textit{scarf} & - & - & 182\\
\cline{2-5}
& \textit{socks} & - & - & 283\\
\cline{2-5}
& \textit{sunglass} & - & - & 150\\
\cline{2-5}
& \textit{tie} & - & - & 134 \\
\cline{2-5}
& \textit{watch} & - & - & 141 \\
\cline{2-5}
& \textbf{Total} & - & - & \textbf{2696}\\
\hline
\multicolumn{2}{c|}{\textbf{Total No. of Items}} & \textbf{2934} & \textbf{7331} & \textbf{22821} \\
\hline
\multicolumn{2}{c|}{\textbf{Total No. of Outfits}} & \textbf{27763} & \textbf{14769} & - \\
\hline
\multicolumn{2}{c|}{\textbf{No. of top-wear, bottom-wear pairs}} & \textbf{6784} & \textbf{8042} & - \\
\hline
\multicolumn{2}{c|}{\textbf{No. of bottom-wear, foot-wear pairs}} & \textbf{20640} & \textbf{8152} & - \\
\hline
\multicolumn{2}{c|}{\textbf{No. of top-wear, foot-wear pairs}} & \textbf{22080} & \textbf{9805} & - \\
\bottomrule
\end{tabular}
\end{table*}
\cbend

\cbstart
The annotation task involves tagging an occasion (e.g., formal or casual) to each combination, followed by rating individual items in the combination (capturing aesthetic value contributed by each item) and an overall rating that determines the suitability of the combination for the mentioned occasion. Fig. \ref{subfig:annotation_desktop}, shows an example of annotating a positive outfit combination, where the presented checked shirt, shaded blue jeans and ankle-boots are all compatible with each other. Here the occasion has been marked as casual; top, bottom and foot are assigned ratings of 5, 3 and 5 stars, respectively, followed by an overall rating of 4 stars for the combination. 

In case of combinations having highly incompatible items, we instruct the annotators to mark one or more specific item types which are a mismatch in the combination, e.g., in Fig. \ref{subfig:annotation_mobile}, the track-pant is not compatible with the checked shirt and the laceup-shoe. So, it has been marked as mismatch and rated with 1 star. Whereas, the shirt and the laceup-shoe are compatible with each other, and rated with 3 and 4 stars, respectively. Here the occasion is tagged as casual and an overall rating of 3 stars is assigned based on the compatibility of top-wear and foot-wear only. We consider this as a negative outfit combination with one incompatible item. Similarly, all of the three items may be incompatible with each other. 

We recruit, five male 
and one female freelance fashionistas with in depth knowledge in the domain of fashion as annotators. We ensure annotation of each set \{$x_{tw}$, $x_{bw}$, $x_{fw}$\} by one female and two male fashion experts. If an item is marked as mismatch by at least two annotators, then we consider that item as incompatible with other two items, otherwise it is compatible. If all of the items are compatible, then the set is considered as a positive outfit, otherwise negative outfit. For simplicity, we label outfits as 0 (negative) or 1 (positive).

At first, we split the (Amazon + Myntra) outfit dataset into two parts $80\%$ (\textit{part 1}) and $20\%$ (\textit{part 2}). 
For getting compatible and incompatible item pairs, we use the outfit combinations from \textit{part 1}.
Each sample combination comprise three items: $x_{tw}$, $x_{bw}$ and $x_{fw}$. Now, if all of the three items are compatible with each other, then 
$(x_{a}, x_{b})\in \mathcal{D}^+_{a,b}$ ($\forall a,b \in \mathcal{T}$, $a\neq b$).
In case, only $x_{bw}$ is incompatible with other two items, but $x_{tw}$ is compatible with $x_{fw}$, then $(x_{tw}, x_{bw})\in \mathcal{D}^-_{tw,bw}$, $(x_{bw}, x_{fw})\in \mathcal{D}^-_{bw,fw}$ and $(x_{tw}, x_{fw})\in \mathcal{D}^+_{tw,fw}$. Similarly, if $x_{tw}$ or $x_{fw}$ is the only incompatible item, we make incompatible pairs accordingly. Finally, if all of the three items are incompatible with each other, then $(x_{a}, x_{b})\in \mathcal{D}^-_{a,b}$ ($\forall a,b \in \mathcal{T}$, $a\neq b$).

After creating $D^+_{a,b}$ and $D^-_{a,b}$ ($\forall a,b \in \mathcal{T}, a\neq b$),
we split the list of generated pairs into $80\%$ (train) and $20\%$ (validation) ratio, which are then used for learning each of the pairwise item compatibility decoders corresponding to types $(a,b)$. 
Outfits from \textit{part 2} are used as a test set for evaluating the outfit compatibility framework.

Finally, for evaluating the quality of recommendations of BOXREC, we collect user preferences for top-wears, bottom-wears and foot-wears, followed by corresponding price-range of each type and an overall shopping budget from $600$ distinct users. 
We use this entire data to evaluate the quality of recommendations provided by BOXREC. For training the PCW framework \cite{Dong:2019:PCW} (used as baseline) we consider 80\% of the user item interactions for training and 20\% for testing.
\cbend


\subsection{Implementation Details} 
We initialize the model parameters of our image encoder, which is based on Google Inception V3 architecture \cite{Szegedy:2016}, with pre-trained weights of ImageNet \cite{imagenet:cvpr09}. It is then fine-tuned using transfer learning \cite{Yosinski:2014} 
on our (Amazon + Myntra + Zalando) clothing item data, classifying 27 item categories (Table \ref{table:dataset_stats}).
We initialized the parameters of each compatibility decoder using the Xavier method \cite{Glorot:2010}.
The convolutional feature map $\mathbf{F}_t \in \mathbb{R}^{D_1\times H_1\times W_1}$ extracted from the fifth Inception C module of Google Inception V3 architecture (Fig. \ref{fig:img_encoder}) has $D_1 = 768$, $H_1 = W_1 = 17$. 
The height and width of $\mathbf{F}_t$ are reshaped as $M = H_1\times W_1 = 289$. 
Normally, a bottom-wear is paired with different top-wears and a foot-wear is used with different top-bottom combinations. So, we set $m_{tw}=15$, $m_{bw}=3$ and $m_{fw}=2$ after checking values in the range $[2, 25]$.

The vocabulary $V$ has cardinality $|V|=920$. We set the size of visual and textual feature vector $A_1 = 900$. Finally, we set
$B_1 = 1024$. Regularization hyper-parameters $\lambda_1$ and $\lambda_2$ are set as $0.00001$ and $0.01$, respectively by grid searching in [0.1, 0.01, 0.001, 0.0001, 0.00001, 0.000001, 0.0000001].
We use AdamW \cite{Loshchilov:2019:AdamW} optimizer, with initial learning rate $\alpha=10^{-3}$, reducing $\alpha$ by a factor of $10$ every $5$ epochs with weight decay $\omega = 0$. 
The default values are kept for two momentum parameters, i.e., $\beta_1=0.9$, $\beta_2=0.999$ and $\epsilon=10^{-8}$. 
We use mini-batch size as $128$ and $32$ for image encoder and compatibility decoder, respectively by grid search and apply gradient clipping \cite{Pascanu:2013} with range [$-5, 5$] (only for compatibility decoder). The parameters of the image encoder is re-trained along with the entire outfit scoring framework, and the implementation is done in PyTorch\footnote{\url{https://pytorch.org/}}. 
\subsection{Evaluation of Outfit Scoring Framework}
We perform a comparative evaluation of BOXREC-OSF (trained on the Amazon + Myntra outfit dataset), with respect to a set of state of the art baselines using several standard metrics. BOXREC-OSF and the corresponding baselines do not consider the price information during the training phase.
\subsubsection{Baseline Methods} The baseline methods are enumerated as follows: 
\begin{itemize}
    \item FashionNet-C: FashionNet-C \cite{He:FashionNet:2018} extracts CNN features (VGG-Medium \cite{Chatfield:2014}) of item images and trains separate matching networks for each item pair types. Finally, matching probabilities of all the item pairs produced by the corresponding matching networks are added together to get a final outfit compatibility score. The learning is modeled as a learning-to-rank framework. We modify the learning as a binary classification to match with our task.
    \item DMLFCAD: DMLFCAD \cite{KedanLi:2019} encodes each fashion item image using ResNet18 \cite{He:ResNet:2016} architecture, and computes three types of joint feature vector representations (dot, diff and sum). The concatenation of these three joint representation is then used to learn a compatibility prediction function to output a compatibility score between item pairs. Outfit compatibility is determined by averaging the scores of all item pairs in an outfit.
    \item NOR-CG: NOR-CG \cite{Lin:2019} uses a two-layer CNN to extract visual features and then applies a mutual attention mechanism to transform visual features to latent representations. The visual features are then concatenated with latent factor \cite{Koren:2009,Lee:2000:ANM,Salakhutdinov:2007} features. Finally, a matching decoder is used to predict the compatibility between tops and bottoms. 
    \cbstart
    \item SCE-Net: SCE-Net  \cite{Tan:2019:LSCWES} mainly focus on learning similarity of fashion items across different aspects without any explicit supervision. i.e., no manually annotated labels like color, shape, pattern, etc. are used to learn the similarity between images of fashion items. Here the number of aspects or conditions is a hyper-parameter.
    \cbend
\end{itemize}
\subsubsection{Evaluation Metrics}
Compatibility AUC \cite{KedanLi:2019,Lin:2019,Vasileva:2018:LTAE} is a standard metric in evaluating outfit scoring frameworks. However, DMLFCAD \cite{KedanLi:2019} emphasizes that Pairwise AUC is a more generalized metric for the evaluation of outfits having a variable number of items. So we make use of Pairwise AUC in addition to Compatibility AUC and Accuracy as evaluation metrics in our outfit score prediction experiments. 

Generally, compatibility scores of item pairs in an outfit are added or averaged for getting the overall outfit score \cite{He:FashionNet:2018, KedanLi:2019}. 
In this paper, we use a new aggregation scheme that performs logical AND operation on all the compatibility scores (Sub-section \ref{outfit_scoring_framework}) of item pairs in an outfit to get an aggregated single score. 

We compute the Pairwise AUC of three pair types, i.e., ($tw,\ bw$), ($bw,\ fw$) and ($tw,\ fw$) and express their average as AP AUC. We denote average and logical AND based aggregation schemes as C\textsubscript{1} and C\textsubscript{2}, respectively. Now, Compatibility AUC corresponding to C\textsubscript{1} and C\textsubscript{2} aggregation are denoted as C\textsubscript{1} AUC and C\textsubscript{2} AUC, respectively. Finally, Accuracy of outfit compatibility is computed based on the C\textsubscript{2} aggregation. 

\subsubsection{Results on (Amazon + Myntra) Outfit Dataset}
The prediction results of compatibility decoders for different pair types and overall outfit compatibility scoring are shown in Table \ref{table:perf_of_sc_pairwise}. From the results in Table \ref{table:perf_of_sc_pairwise}, we make the following observations:
\cbstart
\begin {table*}[t]
\caption {Comparision of BOXREC-OSF with different Baseline Methods.}\label{table:perf_of_sc_pairwise}
\centering
\begin{tabular}{l | c  c  c | c c c c}\toprule
\thead{\multirow{3}{*}{\textbf{Method}}} &
\multicolumn{3}{c|}{\thead{\textbf{Pairwise AUC}}} &
\multicolumn{1}{c}{\multirow{3}{*}{\centering \thead{\textbf{AP} \\ \textbf{AUC}}}} & 
\multicolumn{1}{c}{\multirow{3}{*}{\centering \thead{\textbf{C\textsubscript{1}}\\ \textbf{AUC}}}} &
\multicolumn{1}{c}{\multirow{3}{*}{\centering \thead{\textbf{C\textsubscript{2}}\\ \textbf{AUC}}}} &
\multicolumn{1}{c}{\multirow{3}{*}{\centering \thead{\textbf{Accuracy(\%)}}}}\\
\cline{2-4}
& \thead{\textbf{($tw,\ bw$)}} & \thead{\textbf{($bw,\ fw$)}} & \thead{\textbf{($tw,\ fw$)}} &
& & &\\
\cmidrule{1-8}
FashionNet-C & 0.873 & 0.780 & 0.815 & 0.823 & 0.756 & 0.804 & 80.52\\
DMLFCAD & 0.899 & 0.811 & 0.837 & 0.849 & 0.789 & 0.829 & 82.96\\
NOR-CG & 0.917 & 0.814 & 0.836 & 0.856 & 0.622 & 0.785 & 78.40\\
SCE-Net & 0.894 & 0.846 & 0.694 & 0.811 & 0.453 & 0.527 & 47.47\\
\textbf{BOXREC-OSF} & \textbf{0.943} & \textbf{0.848} & \textbf{0.866} & \textbf{0.886} & \textbf{0.803} & \textbf{0.854} & \textbf{85.45}\\
\bottomrule
\end{tabular}
\end{table*}
\cbend

\cbstart
\begin{itemize}
\item BOXREC-OSF consistently outperforms all the baseline methods on all the metrics. 
NOR-CG uses a shallow image encoder for encoding images which does not capture intricate details of fashion images.
We use an efficient image encoder based on Google Inception V3 (pre-trained on ImageNet data) and fine-tune it on our (Amazon + Myntra + Zalando) fashion item data that helps to capture more visually expressive features. The mutual attention used in BOXREC-OSF is similar to NOR-CG, but with a modified network structure that fully exploits the potential of mutual attention between a pair of images. Apart from that, we use textual features and visual semantic loss in addition to cross entropy and regularization loss to boost the performance of BOXREC-OSF. 
\item NOR-CG has a serious drawback because of using latent factor features of items. So while testing outfit combinations having new item (not used during training), it does not perform well. The Pairwise AUC scores of NOR-CG is better than FashionNet-C and DMLFCAD, but it does not perform well across other metrics. This shows that NOR-CG is not suitable for scoring outfits having more than two items. 
\item SCE-Net shows good Pairwise AUC scores on two occasions corresponding to pair types: ($tw,\ bw$) and ($bw,\ fw$). However, it has the worst C\textsubscript{1} and C\textsubscript{2} AUC scores, followed by Accuracy. This is mainly because of focusing more on learning similarities rather than compatibility. 
\item DMLFCAD performs better than the naive FashionNet-C, because of using discriminative features by applying combination of dot, difference and sum operations to visual features of item pairs.
\item The C\textsubscript{2} AUC measure shows higher values than the corresponding C\textsubscript{1} AUC. According to the C\textsubscript{2} aggregation scheme, an outfit is considered good only if all the items in the outfit are compatible with each other. However, in the case of C\textsubscript{1}, the compatibility scores of all the item pairs are added or averaged. If there are some negative outfit samples, where two items are compatible with each other and the third is not compatible with one of the first two items, then C\textsubscript{1} aggregation will result in high outfit scores. As a result, C\textsubscript{1} predicts those outfits as positive (i.e. false-positive predictions). However, C\textsubscript{2} naturally alleviates such issues, which makes it more robust.
\item Pairwise AUC of $(tw,\ bw)$ is greater than $(bw,\ fw)$ and $(tw,\ fw)$, because the compatibility features of top-wears and bottom-wears are more discriminative. It is easier and more intuitive to check the compatibility between a top-wear and bottom-wear pair than matching bottom-wear with foot-wear or top-wear with footwear.
\end{itemize}
\cbend

\cbstart
\begin {table*}[t]
\caption {Comparison of different variants of BOXREC-OSF}
\label{table:ablation_test}
\centering
\begin{tabular}{l | c  c  c | c c c c}\toprule
\thead{\multirow{3}{*}{\makecell{\textbf{Feature}\\ \textbf{Combinations}}}} &
\multicolumn{3}{c|}{\thead{\textbf{Pairwise AUC}}} &
\multicolumn{1}{c}{\multirow{3}{*}{\centering \thead{\textbf{AP} \\ \textbf{AUC}}}} &
\multicolumn{1}{c}{\multirow{3}{*}{\centering \thead{\textbf{C\textsubscript{1}} \\ \textbf{AUC}}}}
&
\multicolumn{1}{c}{\multirow{3}{*}{\centering \thead{\textbf{C\textsubscript{2}} \\ \textbf{AUC}}}} &
\multicolumn{1}{c}{\multirow{3}{*}{\centering \thead{\textbf{Accuracy(\%)}}}}\\
\cline{2-4}
& \thead{\textbf{($tw,\ bw$)}} & \thead{\textbf{($bw,\ fw$)}} & \thead{\textbf{($tw,\ fw$)}} &
& & \\
\cmidrule{1-8}
$m\text{-}attn$ & 0.942 & 0.841 & 0.856 & 0.879 & 0.797 & 0.852 & 85.25 \\
$[m\text{-}attn\text{,}\ dds]$ & \textbf{0.945} & 0.838 & 0.860 & 0.881 & 0.799 & 0.848 & 84.93  \\
$[m\text{-}attn\text{,}\ txt]$ & 0.944 & 0.844 & 0.865 & 0.884 & 0.802 & 0.853 & 85.37  \\
\textbf{\textit{m-attn + txt}} & 0.943 & \textbf{0.848} & \textbf{0.866} & \textbf{0.886} & \textbf{0.803} & \textbf{0.854} & \textbf{85.45}\\
\bottomrule
\end{tabular}
\end{table*}
\cbend

\cbstart
We also did an ablation study to illustrate the performance of different variants of BOXREC-OSF based on various combinations of features. As shown in Table \ref{table:ablation_test}, with only mutual attention ($m\text{-}attn$), BOXREC-OSF outperforms all of the above-mentioned baselines. 
The combination of $m\textup{-}attn$ and dot, diff, sum ($dds$) operation of DMLFCAD shows better Pairwise AUC scores corresponding to ($tw,\ bw$) and ($tw,\ fw$).
We then test two methods of combining $m\textup{-}attn$ and $txt$ features, e.g., concatenation and summation. 
Results show that BOXREC-OSF equipped with the summation of $m\textup{-}attn$ and $txt$ features performs the best.
\cbend

\subsection{Evaluation of Recommendations provided by BOXREC}\label{eval_reco}
\begin{figure}[t]
	\centering     
	\subfigure[<1K, <1K, <1K]{\label{subfig:nb_itm_oft_rec_pr1}\includegraphics[height=34mm,width=34mm]{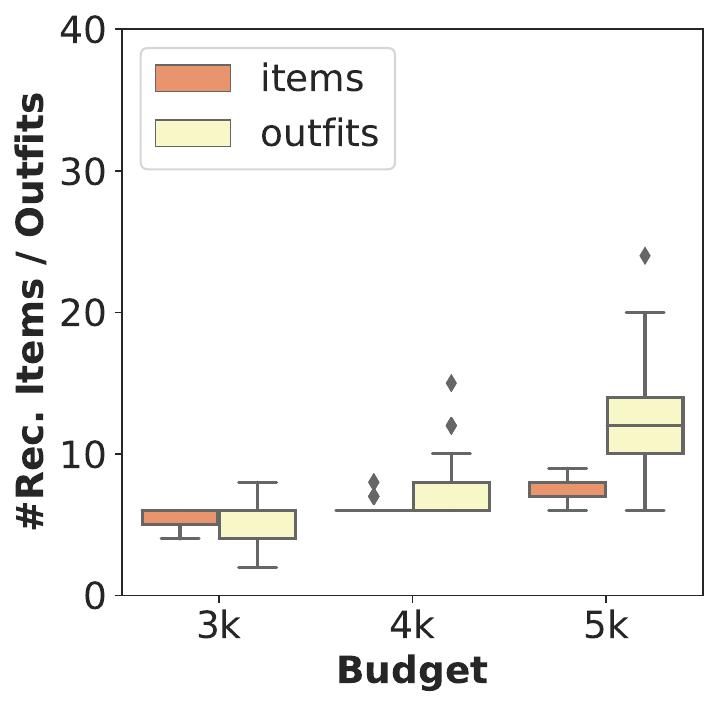}}
	\subfigure[<1K, <1K, 1K-3K]{\label{subfig:nb_itm_oft_rec_pr2}\includegraphics[height=34mm,width=34mm]{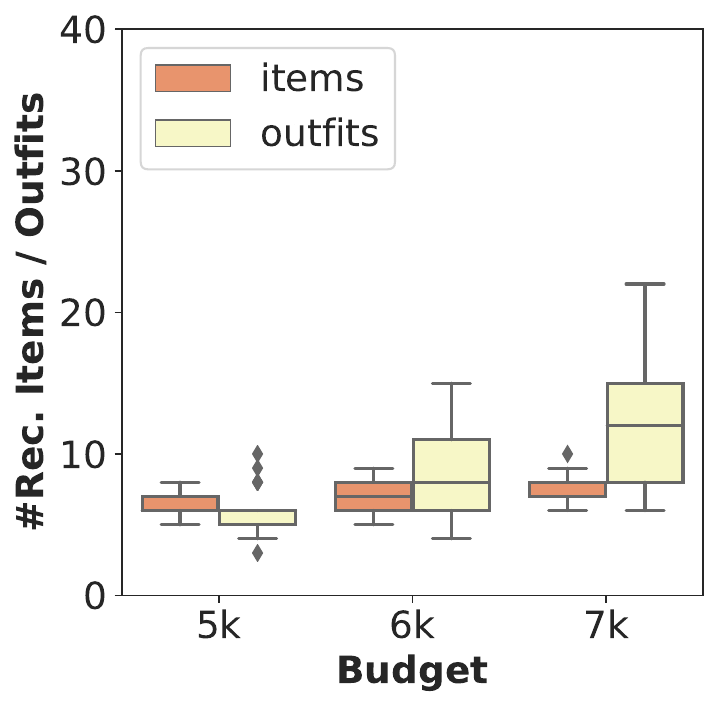}}
	\subfigure[<1K, 1K-3K, <1K]{\label{subfig:nb_itm_oft_rec_pr3}\includegraphics[height=34mm,width=34mm]{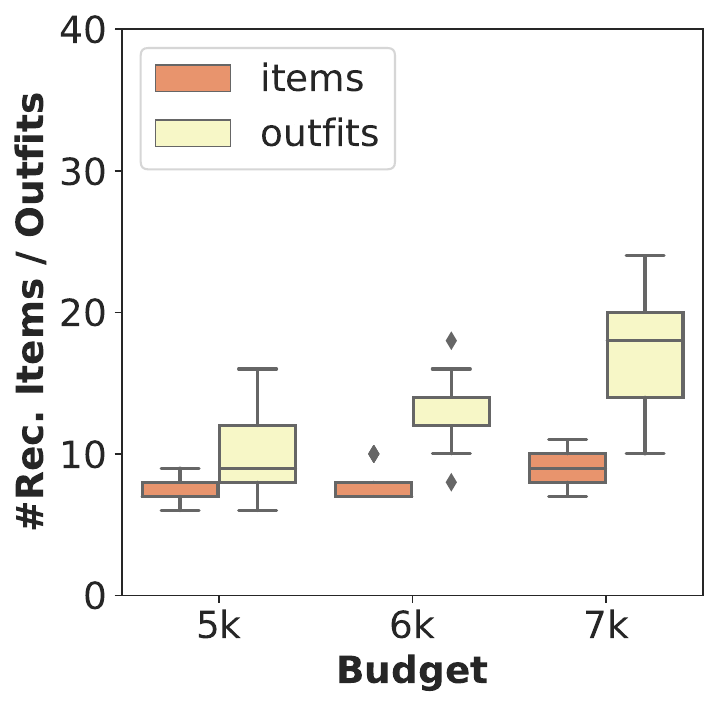}}
	\subfigure[<1K, 1K-3K, 1K-3K]{\label{subfig:nb_itm_oft_rec_pr4}\includegraphics[height=34mm,width=34mm]{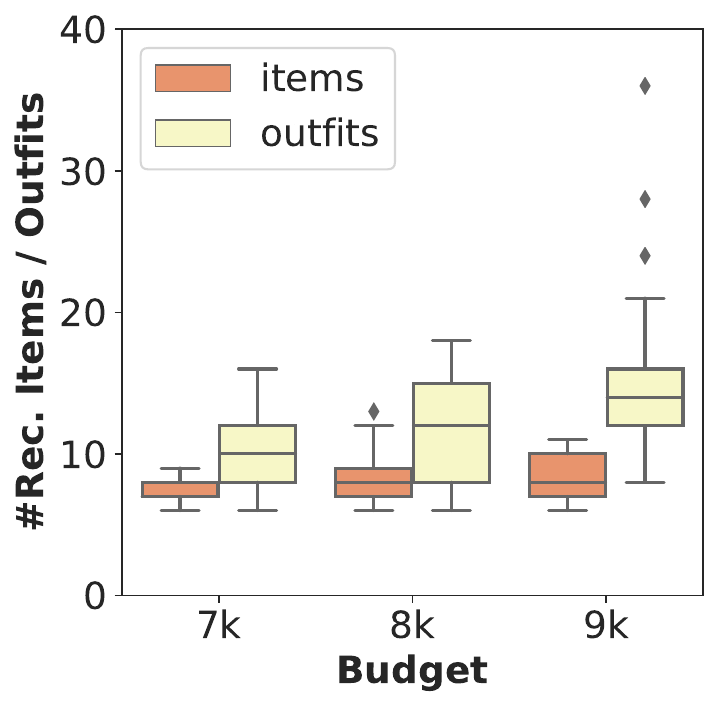}}
	\subfigure[1K-3K, <1K, <1K]{\label{subfig:nb_itm_oft_rec_pr5}\includegraphics[height=34mm,width=34mm]{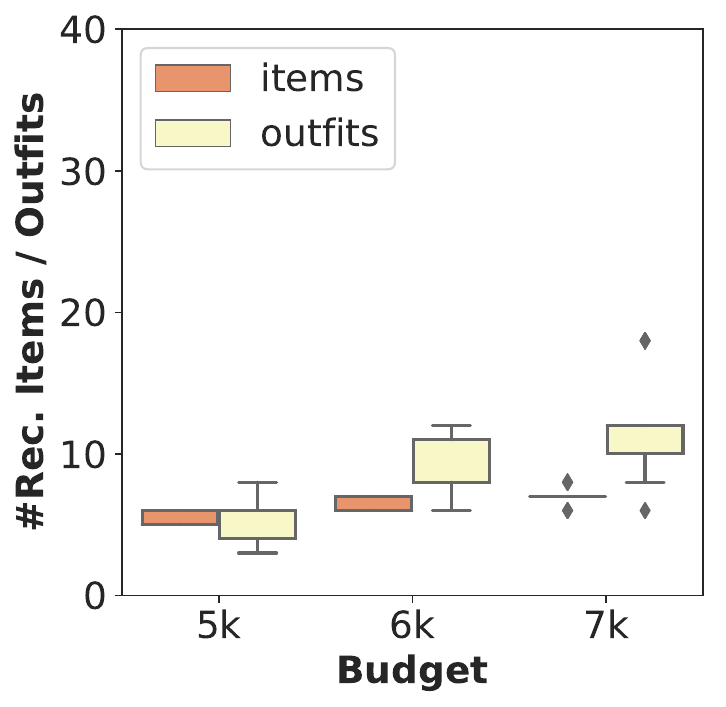}}
	\subfigure[1K-3K, <1K, 1K-3K]{\label{subfig:nb_itm_oft_rec_pr6}\includegraphics[height=34mm,width=34mm]{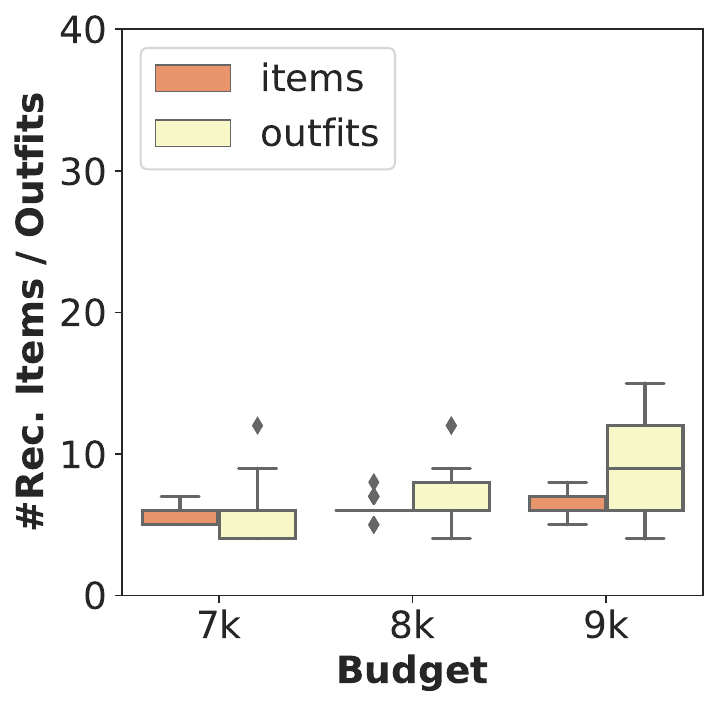}}
	\subfigure[1K-3K, 1K-3K, <1K]{\label{subfig:nb_itm_oft_rec_pr7}\includegraphics[height=34mm,width=34mm]{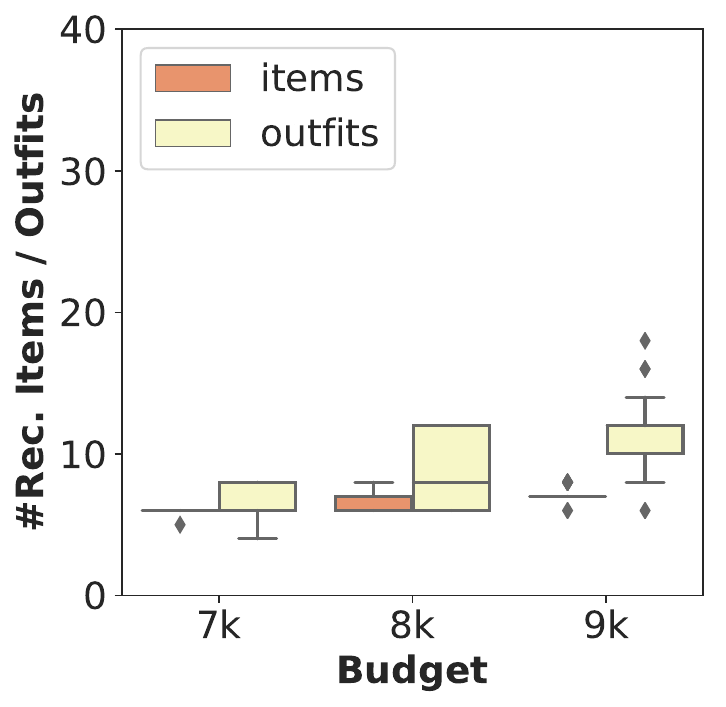}}
	\subfigure[1K-3K, 1K-3K, 1K-3K]{\label{subfig:nb_itm_oft_rec_pr8}\includegraphics[height=34mm,width=34mm]{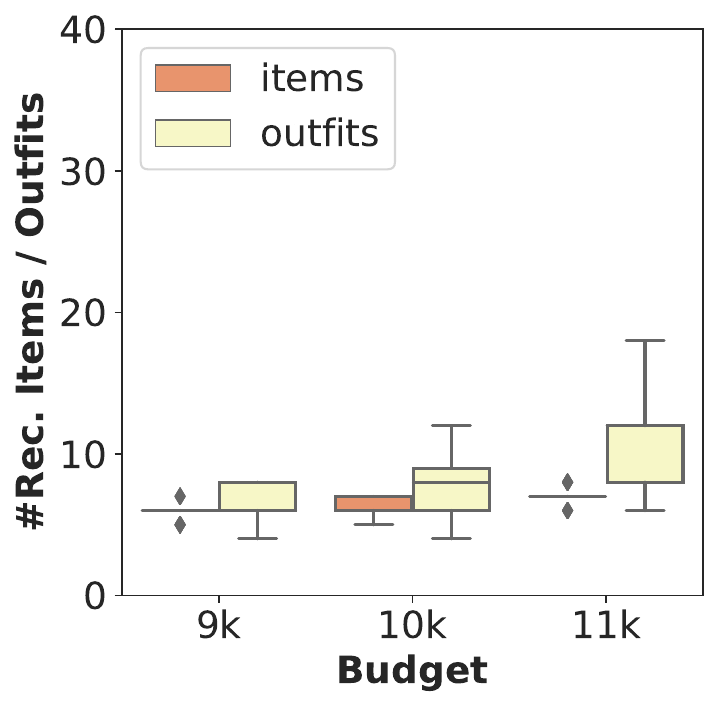}}
	\caption{Number of Items and Outfits recommended by BOXREC for different Item Preferences of Users, across various Type-Specific ($tw$, $bw$, $fw$) Price-Ranges and Overall Budgets (a) to (h).}\label{fig:perf_nb_rec_itm_oft}
\end{figure}

We quantitatively evaluate the recommendations of BOXREC by studying the number of recommended items and corresponding outfits for item preferences of different users across varying type-specific price-ranges and overall budgets, as shown in Fig. \ref{fig:perf_nb_rec_itm_oft}. For qualitative evaluation, we propose a modified version of a standard metric to fit in our setting.
\subsubsection{Evaluation Metrics}
Hit Ratio at top-$n$ ($HR@n$) \cite{Lee:HRATK:2010} is a well known metric for evaluating recommendation systems, which indicates the number of hits found within top-$n$ recommended items. However, BOXREC provides a box full of items within the user-specified budget that maximizes the number of corresponding outfit combinations. Hence, we cannot directly apply $HR@n$ in our context. 
So, we define Hit Ratio ($HR$), a modified version of $HR@n$ that computes the number of hits (for items or outfits) found within a recommended box, as follows:
\begin{equation}
    HR = \frac{\#\text{hits in recommended box}\ H}{\text{total number of products in recommended box}\ H}
\end{equation}
where a product can be either an item or an outfit.
Next, we extend $HR$ to Mean $HR$ ($MHR$), where the mean is taken over all $HR$ values for multiple recommendations, as follows:
\begin{equation}
    MHR = \frac{1}{N}\sum_{j=1}^N HR_j
\end{equation}
where $N$ is the total number of recommended boxes and $HR_j$ is the hit ratio of the box $j$. We use $MHR$ to evaluate the quality of recommendations given by BOXREC both for items and outfits based on user feedback.

\cbstart
\cbend

\subsubsection{Recommendation Results}\label{results:recommendation}
\cbstart
We use two price-ranges: $< 1\text{K}$ and $1\text{K} - 3\text{K}$ for each of the three item types ($tw$, $bw$ and $fw$), which forms a total of eight price-range combinations, with each having three budget values, as shown in Fig. \ref{fig:perf_nb_rec_itm_oft} and Table \ref{table:perf_1_boxrec}. There are $25$ user preferences corresponding to each price-range and budget combination. We also compare BOXREC with PCW \cite{Dong:2019:PCW} (excluding the body shape part). 
\cbend

BOXREC recommends a minimum of $4$ items ($2$ outfits) and a maximum of $6$ items ($8$ outfits) corresponding to price-range combination (<$1$K, <$1$K, <$1$K) and budget $3$K (Fig. \ref{subfig:nb_itm_oft_rec_pr1}). In this case, the average number of recommended outfits is $5$. With increasing budget values ($4$K, $5$K), the number of items and outfits also increases, which is quite intuitive. As shown in Fig. \ref{subfig:nb_itm_oft_rec_pr4}, the maximum number of recommended outfits is more than $35$ for budget $9$K corresponding to price-range combination (<$1$K, $1$K-$3$K, $1$K-$3$K). 
\begin{table*}[t]
\caption {Performance of BOXREC (for formal/casual occasion) based on Average (AVG) Box Price and Mean Hit Ratio~-~$MHR$ of Items and Outfits across different Type-Specific ($tw$, $bw$, $fw$) Price-Ranges and Budgets
(B). All Prices and Budgets are expressed in Thousands (K) (e.g., 1K equals 1000).
}\label{table:perf_1_boxrec}
\centering
\begin{tabular}{c | c  c  c | c | c | c c}
\toprule
\multicolumn{1}{c|}{\multirow{3}{*}{\centering \thead{\textbf{Occasion}}}} &
 \multicolumn{3}{c|}{\centering \thead{\textbf{Price-Range}}} & \multicolumn{1}{c|}{\multirow{3}{*}{\centering \thead{\textbf{B}}}} &
 \multicolumn{1}{c|}{\multirow{3}{*}{\centering \thead{\textbf{AVG} \\ \textbf{Box Price}}}} &
 \multicolumn{2}{c}{\centering \thead{\textbf{MHR}}}
 \\
 \cline{2-4}
 \cline{7-8}
 &
 \thead{\textbf{tw}} &  \thead{\textbf{bw}} & \thead{\textbf{fw}} & & & \thead{\textbf{Item}}& \thead{\textbf{Outfit}}\\
\cmidrule{1-8}
\multicolumn{1}{c|}{\multirow{24}{*}{\rotatebox{90}{formal or casual}}} &
\multirow{3}{*}{$<1\text{K}$} & \multirow{3}{*}{$<1\text{K}$} & \multirow{3}{*}{$<1\text{K}$}
 & 3K & 2.87K & 0.807 & 0.750\\
 & & & & 4K & \textbf{3.90K} & 0.835 & 0.853\\
 & & & & 5K & 4.85K
 & 0.863 & 0.899\\
\cline{2-8}
 &
\multirow{3}{*}{$<1\text{K}$} & \multirow{3}{*}{$<1\text{K}$} & \multirow{3}{*}{$1\text{K}-3\text{K}$} & 5K & 4.89K
& 0.836	& 0.823\\
 &  &  & & 6K & 5.87K
 & 0.852 & 0.864\\
 & & & & 7K & 6.85K
 & 0.865 & 0.908\\
\cline{2-8}
&
\multirow{3}{*}{$<1\text{K}$} & \multirow{3}{*}{$1\text{K}-3\text{K}$} & \multirow{3}{*}{$<1\text{K}$} & 5K & 4.89K & 0.853 & 0.865\\ 
 & & & & 6K & 5.82K
 & 0.868 & 0.916\\
 & & & & 7K & 6.86K
& \textbf{0.881} & \textbf{0.920}\\
\cline{2-8}
&
\multirow{3}{*}{$<1\text{K}$} & \multirow{3}{*}{$1\text{K}-3\text{K}$} & \multirow{3}{*}{$1\text{K}-3\text{K}$} & 7K & 6.87K
 & 0.866 & 0.890\\
 & & & & 8K & 7.83K
 & 0.878 & 0.903\\
 & & & & 9K & 8.84K
& 0.879	& 0.919\\
\cline{2-8}
 &
\multirow{3}{*}{$1\text{K}-3\text{K}$} & \multirow{3}{*}{$<1\text{K}$} & \multirow{3}{*}{$<1\text{K}$} & 5K & 4.71K
 & 0.816 & 0.795\\
 & & & & 6K & 5.69K
 & 0.845 & 0.876\\
 & & & & 7K & 6.70K & 0.857	& 0.902\\
\cline{2-8}
 &
\multirow{3}{*}{$1\text{K}-3\text{K}$} & \multirow{3}{*}{$<1\text{K}$} & \multirow{3}{*}{$1\text{K}-3\text{K}$} & 7K & 6.75K
 & 0.821 & 0.789\\
 & & & & 8K & 7.75K
 & 0.836 & 0.832\\
 & & & & 9K & 8.66K & 0.846 & 0.874\\
\cline{2-8}
 &
\multirow{3}{*}{$1\text{K}-3\text{K}$} & \multirow{3}{*}{$1\text{K}-3\text{K}$} & \multirow{3}{*}{$<1\text{K}$} & 7K & 6.77K & 0.832 & 0.852\\
& & & & 8K & 7.64K
 & 0.848 & 0.878 \\
& & & & 9K & 8.56K
 & 0.860 & 0.907\\
\cline{2-8}
 &
\multirow{3}{*}{$1\text{K}-3\text{K}$} & \multirow{3}{*}{$1\text{K}-3\text{K}$} & \multirow{3}{*}{$1\text{K}-3\text{K}$} & 9K & 8.71K
 & 0.830 & 0.826\\
& & & & 10K & 9.68K
 & 0.842 & 0.842\\
& & & & 11K & 10.63K
& 0.858 & 
0.899\\
\bottomrule
\end{tabular}
\end{table*}

Table \ref{table:perf_1_boxrec} shows the performance of BOXREC based on the average price of the recommended box and $MHR$ of items and outfits provided to various users. 
For each price-range combination, there are three budgets.
The average box price (i.e., mean over 25 recommended boxes of different users) is very close to the overall budget for each price-range combination. The closest being $3.90$K corresponding to the price-range (<$1$K, <$1$K, <$1$K) and a budget of $4$K. For each price-range combination, with an increase in budget, the $MHR$ value of both items and outfits increases. 

The maximum $MHR$ is $0.881$ for items and $0.920$ for outfits, corresponding to the price-range combination of (<$1$K, $1$K-
$3$K, <$1$K) and a budget of $7$K. For price-range combination of (<$1$K, <$1$K, <$1$K) and a budget of $3$K, we got the minimum $MHR$ for items ($0.807$) and outfits ($0.750$). The minimum $MHR$ value of $0.750$ for outfits is mainly because average number of recommended outfits is just above $4$ in this case. So, whenever users do not like a single outfit in such a case, the $MHR$ value drops rapidly. However, the range between $0.750$ and $0.920$ is clearly on the higher side, indicating the superior quality of recommendations.

Since, price-range and budget are not handled by PCW, we compare BOXREC with PCW using $MHR$ of items and outfits. We sample 100 user preferences from the available data for the comparison and collect responses from each user separately for the recommendations provided by PCW and BOXREC. PCW shows an $MHR$ of 0.773 and 0.5699 corresponding to items and outfits, respectively. BOXREC has an $MHR$ of 0.964 and 0.935 for items and outfits, respectively. By analyzing, we found that PCW recommends 9 items to each user that creates 27 possible outfit combinations. But most of the items and outfits are not liked by the user. Thus, BOXREC outperforms PCW.

\begin{figure}[t]
	\centering
	\includegraphics[trim=0 0 0 0,clip,height=90mm]{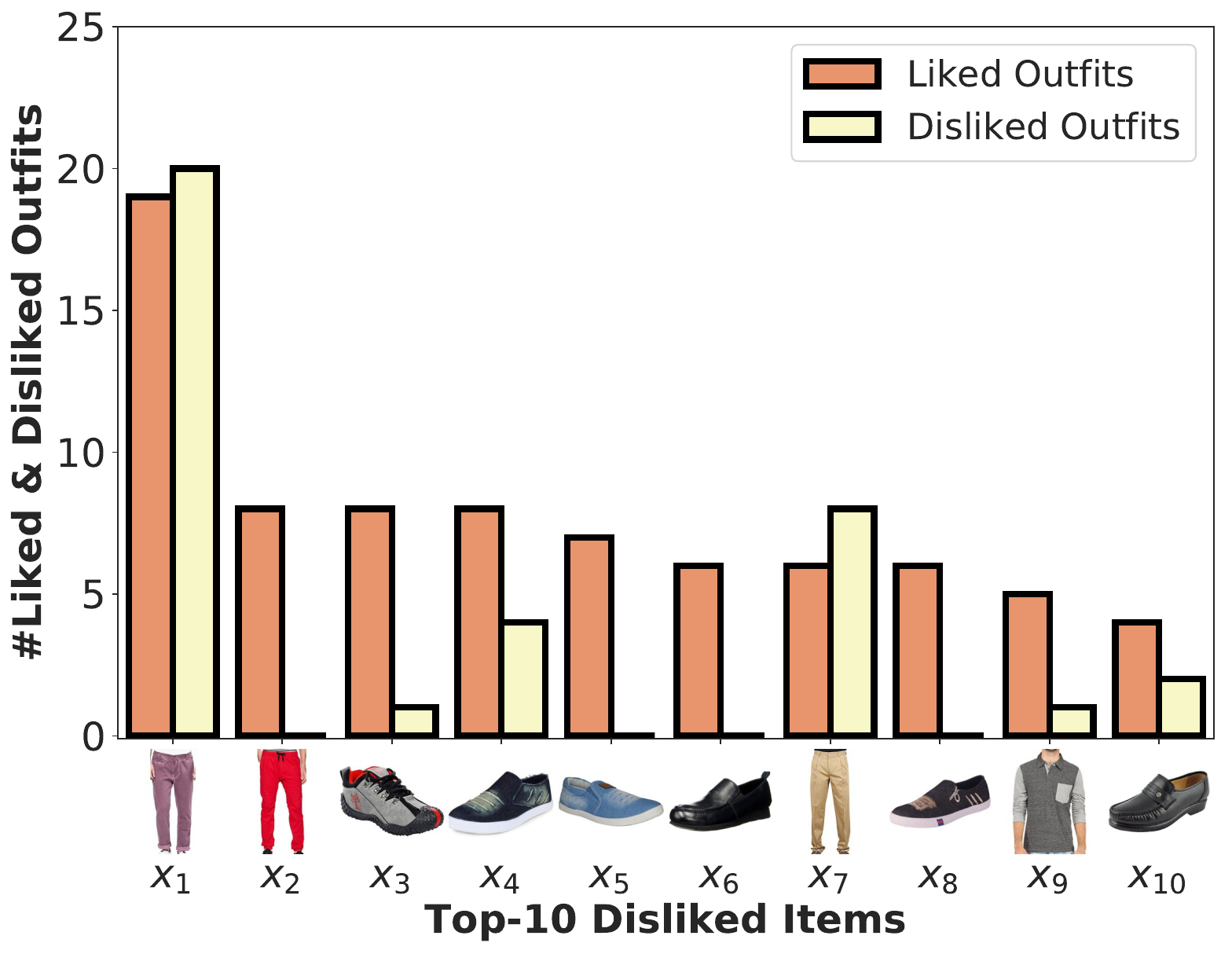}
	\caption{Number of Liked and Disliked Outfits vs Top-10 Disliked Items.
	}
	\label{fig:dl_items}
\end{figure}

We also noted that there are some clothing items that users do not like as a single entity, even though outfits comprised of such items are sometimes liked by them. Fig. \ref{fig:dl_items} shows the number of liked and disliked outfits vs. the top-10 disliked items (sorted in decreasing order of the number of liked outfits). The top-10 disliked items are plotted on the x-axis, and the number of liked and disliked outfits corresponding to each disliked item is plotted on the y-axis. The first item is included in as many as 19 liked outfits and 20 disliked outfits. Moreover, there are some disliked items which are only associated with liked outfits, e.g., the second, fifth, sixth and eighth item. This shows that a disliked item can be part of a compatible outfit combination, when included in the correct ensemble of other items. 
\cbstart
\cbend
\begin{figure}[ht]
	\centering     
	\subfigure[]{\label{subfig:ch_occ}\includegraphics[height=40mm]{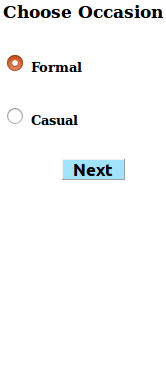}}
	\subfigure[]{\label{subfig:ch_tw}\includegraphics[height=47mm]{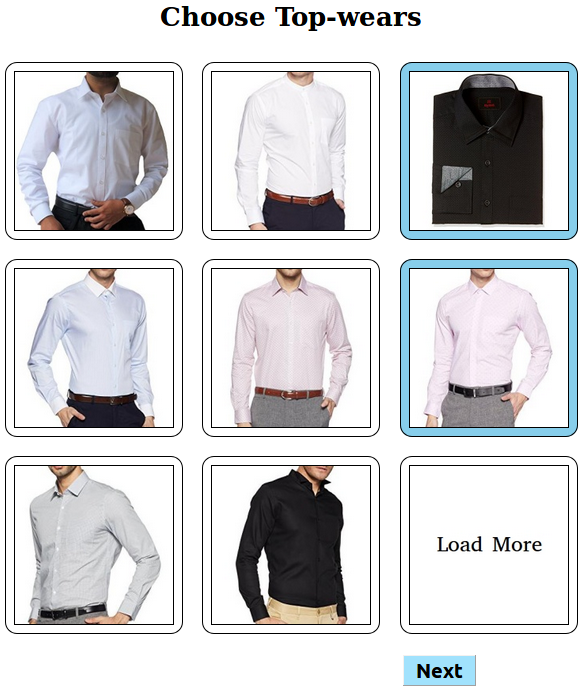}}
	\subfigure[]{\label{subfig:ch_bw}\includegraphics[height=47mm]{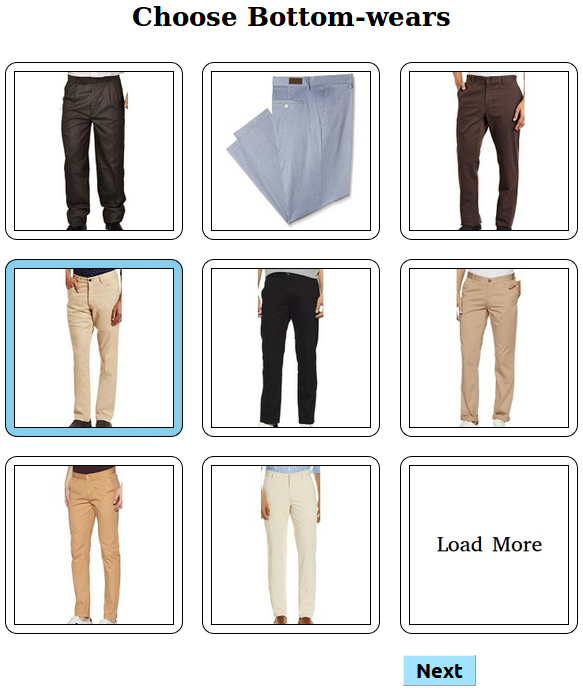}}
	\subfigure[]{\label{subfig:ch_fw}\includegraphics[height=47mm]{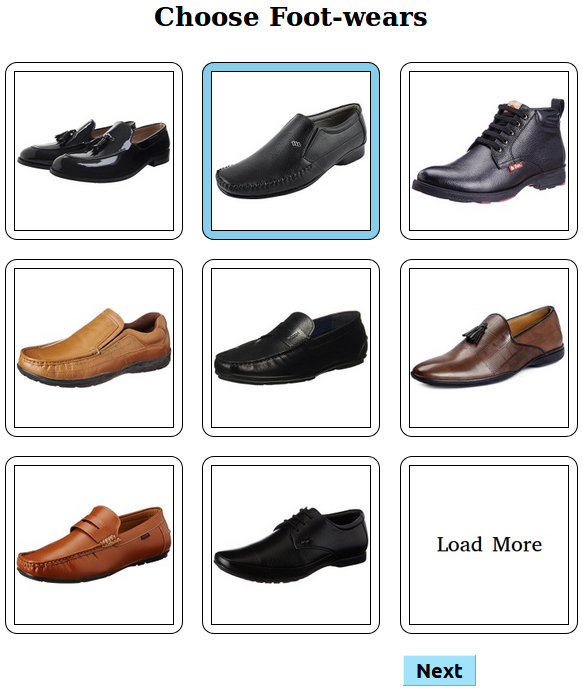}}\\
	\subfigure[]{\label{subfig:pr_budget}\includegraphics[height=60mm]{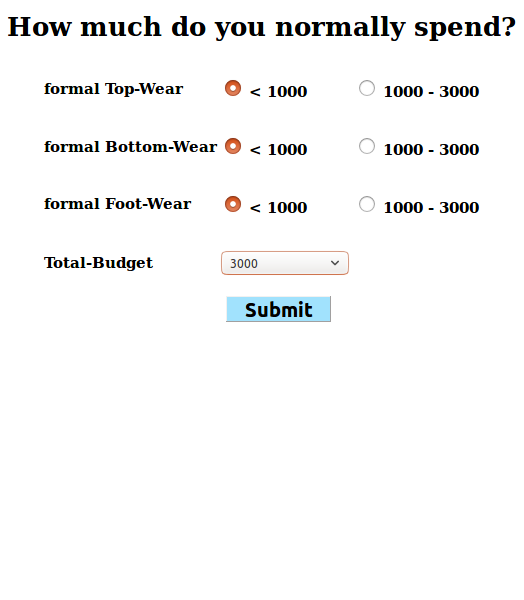}}
	\subfigure[]{\label{subfig:boxrec}\includegraphics[height=80mm]{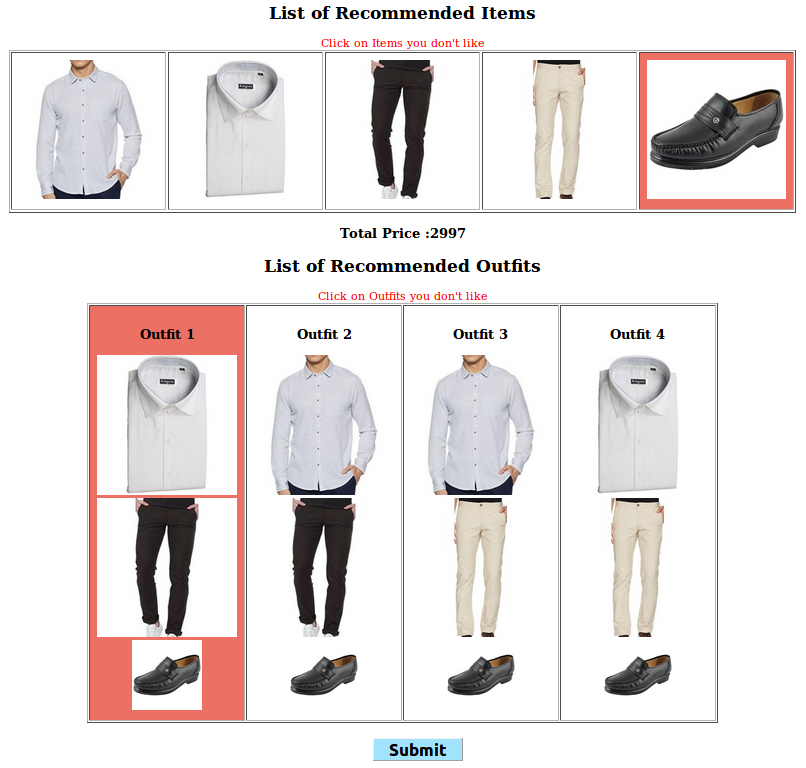}}
	\caption{Implementation of BOXREC as a Web Application. Figures (a) to (e) show how: Occasion (a), Item Preferences (b, c, d), type-specific Price-Ranges and Overall Budget (e) are captured from the user. Finally, a Box containing a list of Items and corresponding Outfits are provided as Recommendation (f).}\label{fig:web_app} 
\end{figure}
\section{BOXREC Web Application}\label{web_app}
We illustrate the implementation of our BOXREC framework as a Web Application using Python Flask in Fig. \ref{fig:web_app}. As shown in Fig. \ref{subfig:ch_occ}, at first, the user is prompted to enter the occasion (i.e., either formal or casual). If the chosen occasion is formal (casual), then a sample of eight formal (casual) top-wear (Fig. \ref{subfig:ch_tw}) items are shown.  The user is then requested to choose the preferred top-wears. If the user does not like any of the shown items, there is an option to load new items. This is repeated for bottom-wear (Fig. \ref{subfig:ch_bw}) and foot-wear (Fig. \ref{subfig:ch_fw}) items as well. 
     
Next, the user needs to select the price-range for individual clothing types and an overall shopping budget (Fig. \ref{subfig:pr_budget}). Finally, BOXREC provides a box filled with items and a list of outfits composed from the given items, satisfying all the preferences and budget constraints. BOXREC also captures the disliked items and outfits for each recommendation (Fig. \ref{subfig:boxrec}), which helps in understanding how disliked items become part of some good looking outfits. A working demo for collecting user's preferences can be found at \url{http://debopriyo.epizy.com/BOXREC/}. 
\section{Conclusion}\label{conclusion}
In this paper, we introduced the BOXREC problem on composite recommendation of fashion items. BOXREC recommends a box full of items to a user such that different combinations of these items maximize the number of outfits suitable for an occasion, while satisfying the user's fashion taste, price-range preference of individual item types and an overall budget constraint. 
\cbstart
BOXREC employs an efficient outfit scoring framework, namely, BOXREC-OSF with $0.854$ C\textsubscript{2} AUC that outperforms the baseline methods.
\cbend
The logical AND based aggregation scheme~-~C\textsubscript{2} is shown to be more robust than the well-known process of adding or averaging over all the pairwise scores~-~C\textsubscript{1}. Based on user feedback, we identified an innovative probable solution for the famous long-tailed problem of existing recommender systems in the fashion domain. We have also demonstrated that disliked or unpopular items can be combined with other items to compose suitable outfits. In the future, we plan to work on the solution for the long-tailed problem in recommendation systems based on our findings about disliked items and make our data and code\footnote{\url{https://github.com/debobanerjee/BOXREC}} publicly available to the research community.
\bibliographystyle{ACM-Reference-Format}
\bibliography{ref}
\end{document}